\documentclass[a4paper]{jpconf}
\usepackage{iopams}
\usepackage{cite}
\usepackage{graphicx}
\begin{document}
\title{Ultrashort pulse action onto thin film on substrate: Qualitative model of shock propagation in substrate explaining phenomenon of fast growth of a hole with increase of absorbed energy}

\author{V~V~Shepelev$^{1}$, N~A~Inogamov$^{2,3}$, V~A~Khokhlov$^{2}$, P~A~Danilov$^{4}$, S~I~Kudryashov$^{4}$, A~A~Kuchmizhak$^{5,6}$ and O~B~Vitrik$^{5,6}$}

\address{$^1$ Institute for Computer-Aided Design of the Russian Aca\-d\-emy of Sciences, Vtoraya Brestskaya 19/18, Moscow 123056, Russia}
\address{$^2$ Landau Institute for Theoretical Physics of the Russian Academy of Sciences, Akademika Semenova 1a, Chernogolovka, Moscow Region 142432, Russia}
\address{$^3$ Dukhov Research Institute of Automatics (VNIIA), Sushchevskaya 22, Moscow 127055, Russia}
\address{$^4$ Lebedev Physical Institute of the Russian Academy of Sciences, Leninsky Avenue 53, Moscow 119991, Russia}
\address{$^5$ Institute of Automation and Control Processes of the Far Eastern Branch of the Russian Academy of Sciences, Radio 5, Vladivostok, Primorsky Krai 690041, Russia}
\address{$^6$ School of Natural Sciences, Far Eastern Federal University, Sukhanova 8, Vladivostok, Primorsky Krai 690041, Russia}

\ead{nailinogamov@gmail.com}

\begin{abstract}
Thin films on substrate are important class of targets for surface nanomodification for plasmonic or sensoric applications. There are many papers devoted to this problem. But all of them are concentrated on dynamics of a film, paying small attention to substrate. In these papers the substrate is just an object absorbing the first shock. Here we present another point of view directed onto dynamics of a substrate. We consider (i) generation of a shock wave (SW) in a support by impact of a contact; (ii) transition from one-dimensional to two-dimensional (2D) propagation of SW; (iii) we analyze lateral propagation of the SW along a film-support contact; and (iv) we calculate pressure in the compressed layer behind the decaying SW. This positive pressure acting from substrate to the film accelerates the film in direction to vacuum. Above some threshold, velocity of accelerated film is enough to separate the film from support. In these cases the circle of separation is significantly wider than the circle of the focal laser spot on film surface. Absorbed laser heat exponentially decays around an irradiated spot $F=F_{\rm c} \exp(-r^2/R_{\rm L}^2)$, where $R_{\rm L}$ is radius of a Gaussian beam, $F$ and $F_{\rm c}$ are local and central fluences, $r$ is a radius from the axis. While the law of decay for the 2D SW in substrate is the power law. Therefore in our case of powerful laser action the edge of a separation circle is defined by propagation of the SW in the support.
\end{abstract}

\section{Introduction}
Laser surface nanostructuring is important direction of studies. It has several subdirections (laser induced periodic surface structures---LIPSS, random structures, solitary creations), which differ in the ways of laser action and in resulting structures.
At relatively weak intensities, wide spots (diameter larger than the comparable wavelengths of light and plasmon-polaritons), optical frequencies (frequency of electromagnetic EM wave should be less than plasmon-polariton frequency), bulk targets (thicker than thickness of a heat affected zone $d_{\rm T}),$ and multiple repetitions the plasmon-polariton mediated structures called LIPSS appears; see recent review \cite{Bonse:LIPSS:2017} and numerous references cited in this review. Usually the LIPSS are in the form of ripples perpendicular to the polarization vector of a EM wave. Their wavelengths are comparable to the optical EM wavelengths (in some cases significantly, few times shorter) and frequency during excitation is equal to laser EM wave frequency.

At larger (than for LIPSS) absorbed energies (above thermomechanical ablation threshold), wide spots (diameter is wider than the scale defined by foaming and surface tension), wide range of EM frequencies (from infrared to soft x-ray; hard x-ray with their large attenuation depth needs separate consideration), bulk targets, single or multiple repetitions, and ultrashort laser pulse durations the random or chaotic surface structures are formed \cite{Vorobyev:06, Fang17}. These structures are formed in the chain of consecutive processes including melting, thermomechanical kick-off, nucleation of cavities, inertial inflation of foam, capillary deceleration of inflation, breaking of foam, and late stage freezing of remnants of cells and membranes of the broken foam.

There are intermediate structures linked to the transition range of energies near the thermomechanical ablation threshold. Porous surface layer made from frozen closed or partially opened cavities is formed in this range of energies \cite{Ashitkov2012}. If central fluence $F_{\rm c}$ of a Gaussian beam is above the ablation threshold then the porous structures form a circle around the main crater. The porous structures are separated from the main crater by the rather high wall; from submicron to few microns height. The wall is a frozen remnant of the shell of the large bubble inflated above the main crater. The shell covers an area where the main crater later forms \cite{Ashitkov2012}. The shell inflates and breaks out leaving the wall where the shell was earlier adjusting (fasten onto) to the bulk target.

The solitary figures called also bumps or cupola like creations appears after laser heating of a small spot. The length of the spot diameter separates the solitary figures from the random structures. Solitary figures appear if the diameter belongs to the range from submicron to few microns long, while the random structures form when the diameter is of the order of ten microns or arbitrarily more.

Usually the solitary figures form in the case of a film target carefully considered below, see next chapters. But solitary figures appearing in a small spot at the surface of a bulk target is also possible. The paper is devoted to the analysis of solitary figures in the case of a thin film supported by substrate.

We investigate the regime of rather large absorbed fluences when the interior of the film inflates after thermomechanical kick-off and flies away leaving a hole in the film. We show that above a definite threshold a diameter of a hole is defined by amplitude of a shock propagating in the substrate under the film.

Shock initiation in substrate by impact of a laser driven plate was considered in the papers \cite{Geraskin:2009,Abrosimov:2014,Krasyuk:2016}. In these papers an approximately planar shock in substrate was driven. A substrate was a layer of finite thickness smaller than diameter of an illuminated spot. The problem of spallation strength of substrate under reflection of a shock from a rear-side of a substrate layer was studied. In the paper below a substrate is thicker than diameter of a spot. Transition from planar to semi-spherical shock shape is described. In the paper \cite{Bolme:film:substrCHemistry:2007} the planar shock driven by an ultrashort laser pulse in a metal film in a liquid substrate was used to initiate fast chemical reactions in a chemically active media (e.g. high explosives).

\section{General view}


\begin{figure}
\centering 
\includegraphics[width=0.76\columnwidth]{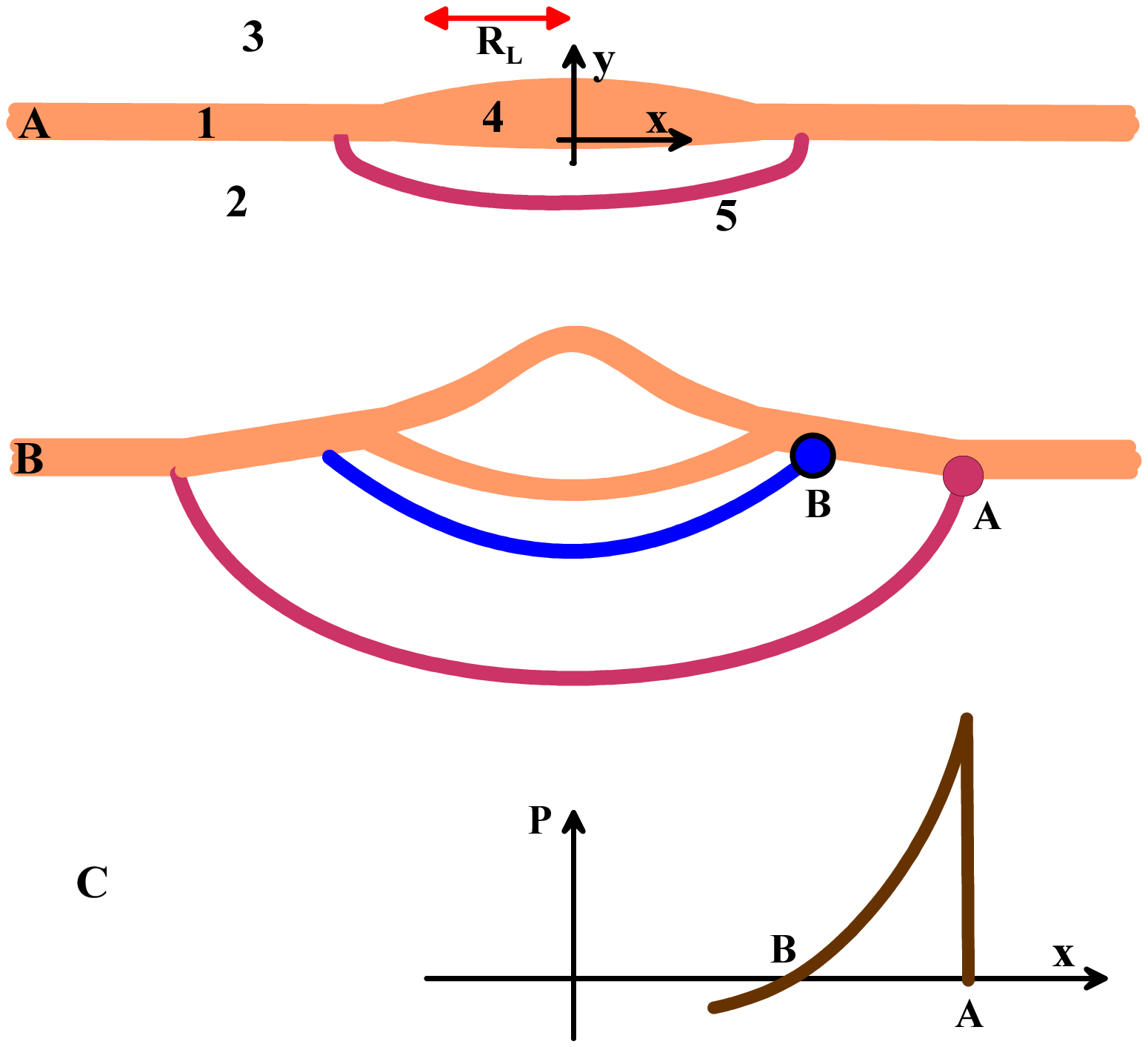}
\caption{Gaussian beam action onto the gold film 1 placed above thick silica substrate 2. The two-side arrow $R_{\rm L}$ is e-folding radius of the laser beam; intensity drops e-times if we deviate to the distance $R_{\rm L}$ from a beam axis. Vacuum 3 is above the film. Part~\pt(a) shows initial bulging 4 of the film inside the illuminated spot and propagation of the shock 5 into glass. Part~\pt(b) presents situation later in time when expansion of the film into the substrate side stops and the region of negative pressures appears. This expanding with time region is located between the bottom of the film and the blue curve. The blue curve is connected with a rarefaction wave propagating in the direction downward from the vacuum--film boundary. Part~\pt(c) shows the instant pressure distribution in the shock compressed layer between the red and blue curves. Namely positive pressure in this layer moves the film up at the interval A--B in part~\pt(b).
 }\label{fig1}
\end{figure}

There are many experimental and theoretical papers devoted to the problem concerning ultrafast laser irradiation of thin films mounted on supporting substrate \cite{Korte:2004,Nakata2003,Nakata:2007,Nakata:2013,Kuznetsov:2009,Nakata:2009,Unger:2012,Meshcheryakov2005,Meshcheryakov2013,Ivanov2013,Ivanov2017,Domke:12,Rouleau2014,SHIH20173,Inogamov2014nanoBump,Inogamov2015nanoBump,Inogamov2016nanoscResLett,AnisimovFLAMN:2017,Wang2017}. Active works on this problem is motivated by important applications in nanoelectronics, plasmonics, sensorics (e.g., SERS---surface enhanced Raman scattering), and fabrication of metasurfaces \cite{Wang2017,Chen2014,Kuchmizhak2015,Zywietz2014,Kuchmizhak2016nanoscale, Kuchmizhak:15}. Laser action transfers initially smooth film into nanostructured one \cite{Korte:2004,Nakata2003,Nakata:2007,Nakata:2013,Kuznetsov:2009,Nakata:2009,Unger:2012,Meshcheryakov2005,Meshcheryakov2013,Ivanov2013,Ivanov2017,Domke:12,SHIH20173,Inogamov2014nanoBump,Inogamov2015nanoBump,Inogamov2016nanoscResLett,AnisimovFLAMN:2017,Wang2017,Kuchmizhak2016nanoscale}.

Films thinner than thickness of a heat affected zone $d_{\rm T}$ are called thin here (for gold $d_{\rm T}\approx 150$~nm). Their dynamics, freezing at late times, and structures fixed during the freezing processes are qualitatively different from those which correspond to the bulk targets; one-dimensional (1D) dynamics of bulk targets was studied in papers \cite{POVARNITSYN20129480,Ashitkov-Ta-2015,ROMASHEVSKIY201612,Povarnitsyn-arxiv-2014,Stegailov-Zhilyaev-Pe:2015,POVARNITSYN20151150,Norman2012,Ashitkov2018-Ta,Povarnitsyn-PRB:2015}. The thin films are considered in the presented paper.
Experimental papers \cite{Korte:2004,Nakata2003,Nakata:2007,Nakata:2013,Kuznetsov:2009,Nakata:2009,Unger:2012,Ivanov2013,Domke:12,Wang2017} devoted to the film-substrate problem can not follow events in substrate. While theoretical and simulation papers \cite{Meshcheryakov2005,Meshcheryakov2013,Ivanov2013,Ivanov2017,SHIH20173,Inogamov2014nanoBump,Inogamov2015nanoBump,Inogamov2016nanoscResLett,AnisimovFLAMN:2017,Wang2017,Kuchmizhak2016nanoscale} are oriented to exploration of a fate of a film. Thus they try to limit numerical resources spent to description of substrate; indeed, in simulations the resources are restricted. Effective approach excluding substrate from molecular dynamics (MD) simulation was proposed in \cite{Inogamov2014nanoBump,Inogamov2015nanoBump,Inogamov2016nanoscResLett}.

If we are interested in description of separation of a thin film from substrate then we can use important simplifying assumption that film-substrate interaction is very limited in time: it lasts during film acoustic time scale $\sim t_{\rm s}=d_{\rm f}/c_{\rm s}\sim 10$~ps, while evolution of a film {\it after} its separation from substrate is the 3D motion and continues up to nanoseconds. This allows to follow the film-substrate interaction by relatively (relative to MD) fast 1D two-temperature hydrodynamics (1D 2T-HD) \cite{Inogamov2014nanoBump,Inogamov2015nanoBump,Inogamov2016nanoscResLett} and after separation to forget about substrate.

The simplifying assumption that the film-substrate interaction is accurately described in the 1D approximation
is valid over a range of absorbed energies from $F_{\rm abs}=0$ to $F_{\rm abs} < F_{\rm 1D-3D}$.
Above this range we have to pay much more attention to hydrodynamics of substrate because the 3D effects become very important.
This is shown below. The scheme of our case is presented in figure~1.
The scheme helps us to understand how internal motion inside the substrate under ceiling (with a film serving as a ceiling)
influences motion of a film and causes separation of a film {\it outside} the laser heated spot.

The shock going in a substrate is generated by expansion of a fast heated film into the substrate:
a film is heated faster than a sound wave crosses the film: $t_{\rm heating}\ll t_{\rm s}=d_{\rm f}/c_{\rm s};$
the film is heated homogeneously across its thickness
(but not across the lateral extension $\sim R_{\rm L}$ of a laser heated spot) because $d_{\rm f} \ll d_{\rm T}.$
Thus the film-substrate contact in figure~1 acts as a piston driving a shock in substrate.
This driving continues during the time interval with duration of the order of acoustic time scale $t_{\rm s},$
see \cite{Inogamov2014nanoBump,Inogamov2015nanoBump,Inogamov2016nanoscResLett}.

After this interval of pressing, the rarefaction from the vacuum side of a film arrives to the contact
and the piston stops its pressing onto substrate.
Therefore the shock in substrate transfers into so called triangular shock---shock compression of matter
decreases behind the shock front and shock decays during its propagation.
Transition from 1D to 3D expansion of shock in substrate takes place
when the distance that is run by the shock becomes of the order of radius of a spot $R_{\rm L}.$
Below the transition fluence $F_{\rm 1D-3D}$ the shock is weak at the propagation distances $\sim R_{\rm L}.$
Thus it can not affect motion of a film, it cannot induce separation of a film outside the spot $R_{\rm L}.$

But above the transition fluence $F_{\rm 1D-3D}$ the shock is strong enough to cause separation of a film even outside the spot $R_{\rm L}.$

Thin gold film is mounted on a thick glass substrate.
Thickness of the film is $d_{\rm f}=50$~nm. In our simulations we consider two-dimensional (2D) plane motion, our frame is $x,y,$ where the horizontal axis $x$ is going along the initial contact line between the film and the substrate below the film, see figure~1. The vertical line $x=0$ is the axis $y$ and the axis of a laser beam irradiating a film. Duration of a laser pulse $\tau_{\rm L}$ is much shorter than acoustic time scale $t_{\rm s}.$
Therefore we start from the temperature distribution where increase of temperature from its initial value is proportional to a Gaussian function $F=F_{\rm c} \exp(-x^2/R_{\rm L}^2)$ giving distribution of absorbed energy along a gold film. Film is thin, thinner than thickness $d_{\rm T}$ of heat affected zone for metal choosen for a film; as was said $d_{\rm T}\approx 150$~nm for Au; for the thick (500 nm Ag \cite{Kuchmizhak:2017-500nmAg}, \cite{Li2017} 50~ps action onto different molten thick films) films with thicknesses $d_{\rm f} > d_{\rm T}$ the laser caused flow
is very different. In the thin films $d_{\rm f}\ll d_{\rm T}$ the almost constant temperature distribution along the normal direction is established in a few picoseconds; but a lateral heat transfer along the spot lasts during the nanosecond time scale for radii $R_{\rm L}\sim 1$~micron \cite{Inogamov2014nanoBump,Inogamov2015nanoBump,Inogamov2016nanoscResLett}; $R_{\rm L}\gg d_{\rm f};$ we neglect thermal conductivity in dielectric substrate at the considered time scales. The scale $R_{\rm L}$ is shown in figure~1\pt(a). We have in mind experiments similar to that described in \cite{Wang2017}; our new experiments are described in next Section. In our simulations $R_{\rm L} = 0.25$~$\mu$m, and a pulse has subpicosecond duration. Let us mention that in our hydrodynamics simulation it is significant that the simulation box size is larger than the radius $R_{\rm L}.$



\begin{figure}
\begin{center}
\begin{minipage}[b]{0.47\columnwidth}
\includegraphics[width=1\columnwidth]{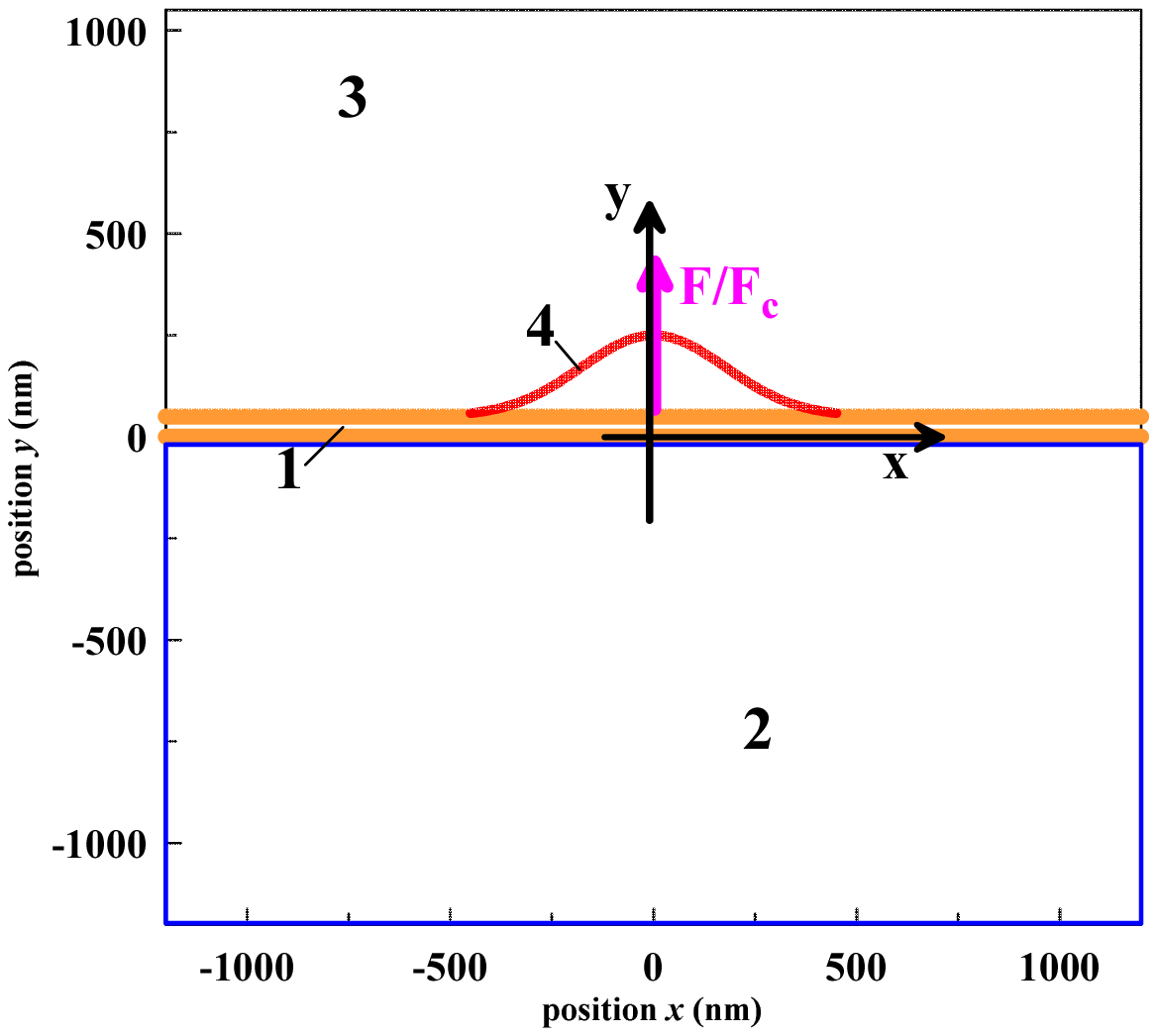}
\begin{center}\pt(a)\end{center}
\end{minipage}
\hspace{0.04\columnwidth}
\begin{minipage}[b]{0.47\columnwidth}
\includegraphics[width=1\columnwidth]{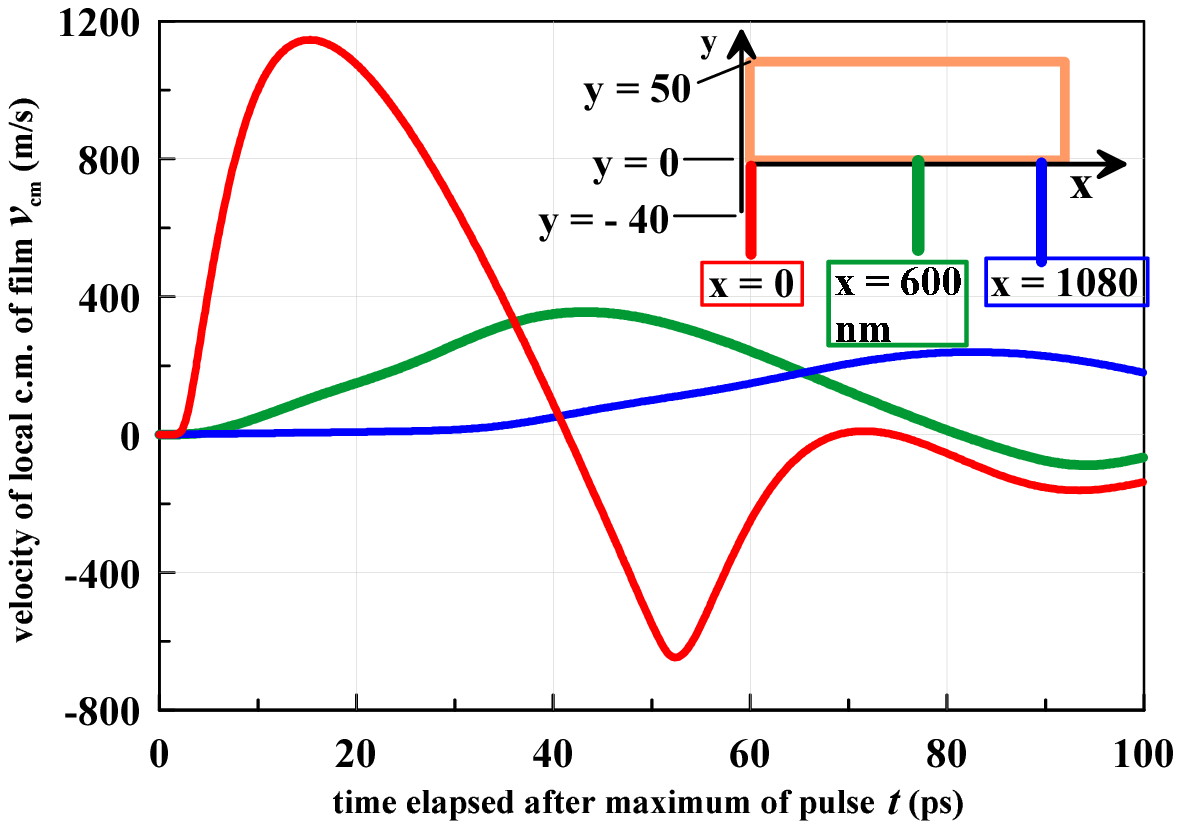}
\begin{center}\pt(b)\end{center}
\end{minipage}
\end{center}
\caption {\label{fig2} Part~\pt(a) shows the simulation box. Horizontal and vertical spatial scales are equal to each other. Initial position of the film is marked by the digit 1, the digit 2 is the silica substrate, 3 means vacuum. The frame $x,y$ is described in text. Gaussian distribution $F/F_{\rm c}$ shown by the curve 4 has e-folding radius $R_{\rm L}=250$~nm. The horizontal distance from the beam axis to the lateral wall of the simulation box is 1200 nm${}= 1.2$~$\mu$m. This distance is almost five times larger than the radius $R_{\rm L}$
(then the ratio $F_{\rm c}/F_{1200}=\exp(x_{1200}^2/R_{\rm L}^2)$ is huge: $10^{10}).$ Thus there is enough space for expansion and decay of the shock in the supporting half-space~2.
Part~\pt(b) shows velocities
$v_{\rm cm}(t) = \int^t (p-p_{\rm ini}) \, \rmd\tau/(\rho_{\rm f} d_{\rm f}),$
see equation~(\ref{eq0:vSW}),
calculated with temporal pressure dependencies $p(x_i, y=-40~{\rm nm}, t)$ taken from simulation presented below. Here the points where the temporal evolution of pressure was followed have the spatial coordinates equal to $x_i,$ $y=-40$~nm, where $i=1, 2, 3.$
The three points with $x_1=0$, $x_2=600$~nm, $x_3=1080$~nm are shown in the inset.
The inset corresponds to the part of the geometrical plane shown as whole in part~\pt(a).
The rising part of dependencies $v_{\rm cm}(t)$ for $i=2, 3$ corresponds to acceleration of the film by pressure in shock compressed underlying silica. The shock compressed area has a shape of a {\it running} half-spherical shell, see figure~1\pt(b).
The colors of the points in the inset link the points to the same colors of the corresponding velocity curves.}
\end{figure}

Initial stage of expansion of a film in its heated region is shown in figure~1\pt(a). Shortly after subpicosecond pulse (at $t\sim 10$~ps) approximately one-dimensional motion takes place. This means that pressure behind the shock wave (SW) (red curve in figure~1\pt(a)) is proportional to local absorbed fluence; indeed $R_{\rm L}$ is significantly larger than thickness $d_{\rm f}$ of a film. Pressure between the SW and the film-glass contact is positive. Later in time the rarefaction wave propagating from the film-vacuum boundary crosses Au-glass (film is made from Au---gold) contact, stops motion of the Au-glass contact in the downward direction, and decrease the pressures above the blue curve in figure~1\pt(b) to negative values. Pressure field in the moving region inside the silica substrate is shown in figure~1\pt(c).

Figure~1\pt(b) shows expansion of the SW in glass substrate under the gold film. SW front shown as the red curve in figures~1\pt(a) and 1\pt(b) moves down and along the film.
The point A in figure~1\pt(a) is the point where SW begins to interact with the film. Pressure is positive in the compressed layer of glass between the blue and red curves in figure~1\pt(b). Compressed layer forms half-spherical shell. Pressure in the shell accelerates film up in direction to vacuum at the interval A--B in figure~1\pt(b). Accelerations acts during finite time interval. Let us consider the observation point A. The SW achieves it at the instant shown in figure~1\pt(b). After that during the time interval needed for the compressed layer to pass by the observation point this point accumulates vertical component of momentum
\begin{equation} \label{eq0:vSW}
\rho_{\rm Au}(t) d_{\rm f}(t) v_{\rm cm}(t) = \int_{-\infty}^t [p(\tau)-p_{\rm ini}] \rmd\tau
\end{equation}
because difference of pressures between vacuum semi space and compressed layer exists and resulting force is directed into vacuum side; here $\rho_{\rm Au}(t)$ and $d_{\rm f}(t)$ are current average density and thickness of the gold film; the product $\rho_{\rm Au}(t) d_{\rm f}(t)\approx const$ due to conservation of mass; local (along $x)$ center of mass of a film is denoted---cm.




Let $t_{\rm B}$ is the instant when the whole compression layer shown in figure~1\pt(b) rolled past the observation point. If velocity
\begin{equation} \label{eq1:vSW}
v(t_{\rm B}) = \left(\int_{-\infty}^{\rm tB} [p(\tau)-p_{\rm ini}] \rmd\tau \right)/(\rho_{\rm Au} d_{\rm f})
\end{equation}
accumulated during the shock acceleration stage will be high enough then the film in the observation point separates from substrate.
The same arguments are used in \cite{Inogamov2014nanoBump,Inogamov2015nanoBump,Inogamov2016nanoscResLett} to explain separation of a film as whole thanks to pressure in substrate accelerating local (along a film) center of mass of a film in the thermomechanical mechanism of separation. After separation pressure field in substrate ceases its influence on the film in the observation point. Therefore the separated film after separation continues its flight in inertial state. Much later, when curvature of a film becomes significant, capillary forces begin to affect motion. At the early stage a film is almost flat.

Let $x_{\rm sw}$ is the end of the interval along a film inside which accumulated velocity $v(t_B)$ is above the threshold for separation from substrate. There are two different parts of bulging of the film if the edge of separation $x_{\rm sw}$ is significantly above the hot spot radius $R_{\rm L}.$ The inner part is defined by the radius $R_{\rm L}$: $x_{\rm tm}\sim R_{\rm L}$ for large enough absorbed fluences $F_{\rm abs}.$ The inner part is the thermomechanically driven part. While the outer part is located between the inner part $x_{\rm tm}$ and the edge $x_{\rm sw}.$ The outer part is the SW driven part. Rather large values of $F_{\rm abs}$ above crossover $F_{\rm abs}^*$ are necessary to achieve the condition $x_{\rm sw} > x_{\rm tm}.$ The inner part in figure~1\pt(b) is splintered to two subfilms (see \cite[figures~7 and~9]{Wang2017} illustrating splintering of a film above threshold for thermomechanical splintering), while in the A--B interval (moved up by SW) the film moves as whole.

For Gaussian beam $F=F_{\rm c} \exp(-x^2/R_{\rm L}^2)$ in the thermomechanical regime the edge of the separated part grows as $$x_{\rm tm} = R_{\rm L} \sqrt{\ln({F_{\rm c}/F_{\rm tm}})},$$ where $F_{\rm tm}$ is threshold for thermomechanical (tm) separation; expression $x_{\rm tm} \propto \sqrt{\ln}$ simply follows from inversion of exponent in Gaussian function. We see that laser fluence exponentially decays with deviation $x$ from the beam axis, therefore the distance $x_{\rm tm}(F_{\rm c})$ slowly (only logarithmically) increases at large fluences. While pressure $p$ in the compressed silica in figure~1\pt(b) and accumulated velocity $v(t_{\rm B})$ (\ref{eq1:vSW}) decay with distance $x$ more slowly---they $(p$ and $v)$ obey power law decrease with distance. Thus above the crossover $F_{\rm abs}^*$ the separation process at the edge will be led by SW in silica.

If we plot separation distance $x_{\rm sep}$ as function of central fluence $F_{\rm c}$ of the Gaussian beam then for $F < F^*$ we will have logarithmic increase $$\propto \sqrt{\ln({F_{\rm c}/F_{\rm tm}})}$$ while above the crossover $F^*$ the distance $x_{\rm sep}$ begins to increase more quickly. This explains existence of the kink in the dependence $x_{\rm sep}(F)$ observed in \cite{Wang2017} in figure~1(h) in this paper and in the new experimental data in next Section.

Figure~2 presents the largest simulation box in the 15 different runs. We impose free streaming boundary conditions at all four walls of the rectangular box to avoid reflection of SW. Absorbed fluence $F_{\rm abs}$ increases temperature homogeneously along thickness of the film $0<y<50$~nm and proportionally to Gaussian distribution along the axis $x.$ We use one-temperature (1T) 2D hydrodynamic code (1T-2D-HD) thus neglecting fine peculiarities linked to the two-temperature (2T) state. These peculiarities were studied in papers \cite{Inogamov2014nanoBump,Inogamov2015nanoBump,Inogamov2016nanoscResLett,INA-nanoBump-J.Ph.Con.Ser:2016,Inogamov2016APA}. For bulk gold a 2T stage has been analyzed in \cite{Ashitkov-Au-bulk-eps:2016J.Ph.Conf.Ser}.

In \cite{Shepelev2D-2T-1T.heat.expansion.J.Ph.Conf.Ser:2018} we numerically have studied effects of 2D spreading of absorbed heat from a laser focal spot along thickness and length of a film, see also \cite[chapter~3.2]{INA-nanoBump-J.Ph.Con.Ser:2016} for description of analytical approach. Two-temperature stage gradually transforming into one-temperature stage and up to re-crystallization was considered (2D-2T code without hydrodynamics \cite{Shepelev2D-2T-1T.heat.expansion.J.Ph.Conf.Ser:2018}). The fastest thermal spread takes place during 2T stage. Hydrodynamic motion was excluded in \cite{Shepelev2D-2T-1T.heat.expansion.J.Ph.Conf.Ser:2018}. It was shown in \cite{Shepelev2D-2T-1T.heat.expansion.J.Ph.Conf.Ser:2018} that spreading of absorbed heat from a hot laser spot with beam radius $R_{\rm L}$ (during acoustic time scale $t_{\rm s}=d_{\rm f}/c_{\rm s})$ is significant only for small radiuses $R_{\rm L},$ smaller than 150~nm.

Here, in this paper, we come interested in the larger spots. Therefore our 1T-2D-HD code treats dynamics shown in figure~1 in adiabatic approximation (heat conduction is neglected). For small spots $R_{\rm L}<150$~nm the beam e-folding radius $R_{\rm L}$ must be changed to the radius $R_{\rm spr}$ of the thermally expanded spot (spr---spreading). The spot $R_{\rm spr}$ is the hot spot formed at the end of acoustic stage lasting $\approx t_{\rm s}.$ The radius $R_{\rm spr}$ is the e-folding radius for temperature distribution formed at the end of acoustic stage. It changes the radius $R_{\rm L}$ in the simulations describing flight of a film after its separation in the spot \cite{Inogamov2014nanoBump,Inogamov2015nanoBump,Inogamov2016nanoscResLett,freeStandng.J.Ph.Conf.Ser:2018} because the thermomechanical separation process covers only the acoustic stage.
Let us mention that the codes 2D-2T without hydrodynamics and 1T-2D-HD accomplish each other.








\begin{figure}
\centering\includegraphics[width=1\columnwidth]{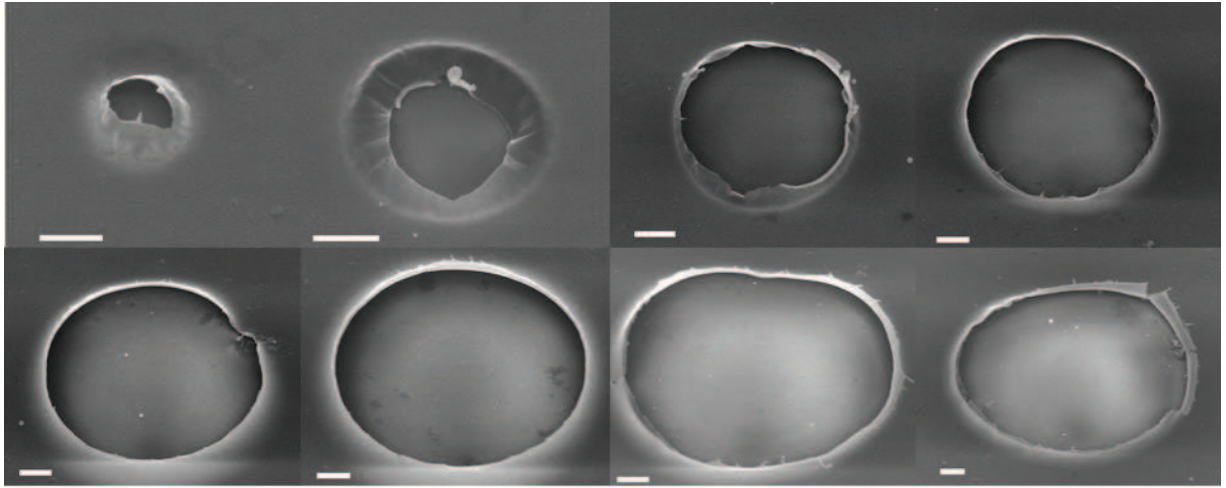}
\caption{
SEM visualization of the damage created after single laser shot are shown.
Above some threshold holes in a film appears, see figure~1 in paper \cite{Wang2017}.
All eight scale bars have length 1 micron.
Energies of pulses which fabricate these nanostructures are equal to
48, 64, 100, 240, 320, 400, 800, and 960 nJ. Radius of the holes grows as energy increases. Wavelength of laser used is 515 nm, its duration FWHM is 200 fs,
This is the 50 nm thick silver film deposited onto a glass substrate.
\label{fig1exprmnt}
}
\end{figure}

\begin{figure}
\centering\includegraphics[width=0.76\columnwidth]{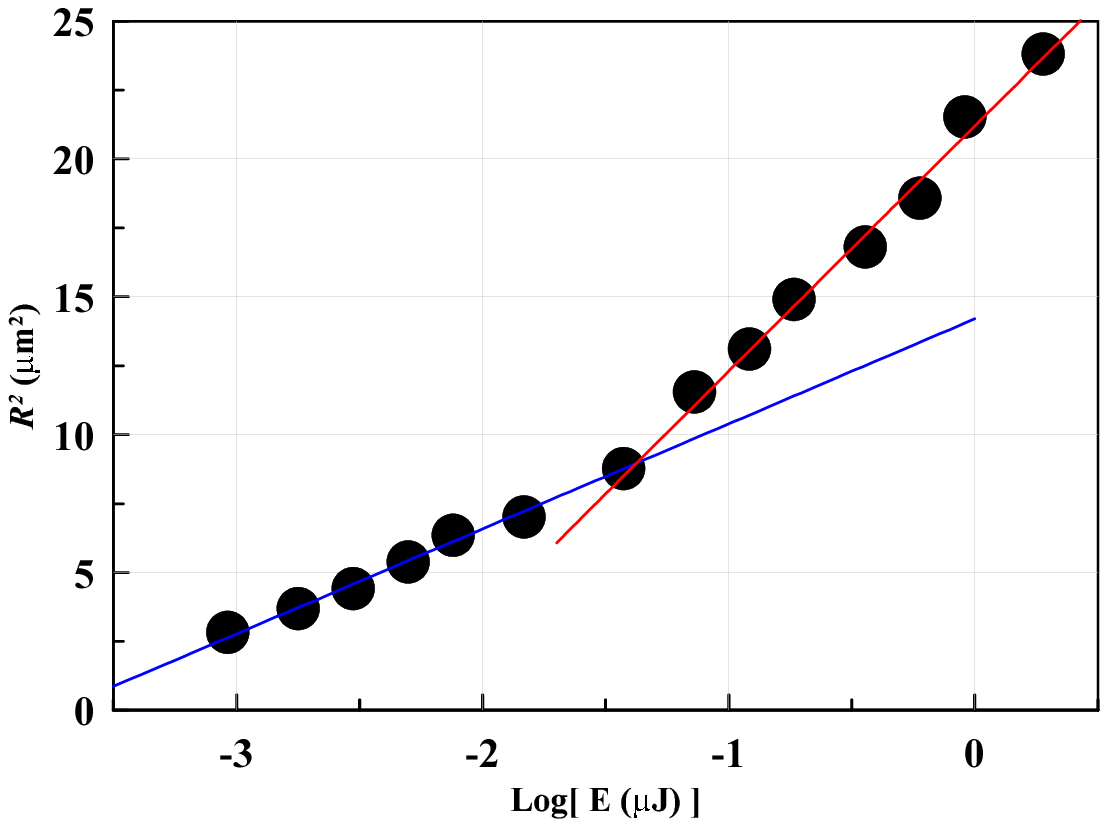}
\caption{
Dependence of square of a radius of the holes shown in figure~\ref{fig1exprmnt} from natural logarithm of energy of a laser pulse which creates the particular hole. The first family of the points relates to the rather weak laser shots with rather small energies of a pulse when the radius grows according to the expression $x_{\rm tm} = R_{\rm L} \sqrt{\ln({F_{\rm c}/F_{\rm tm}})}$ which follows directly from a Gaussian laser energy spatial distribution in a focal spot, see previous Section below the equation~(\ref{eq1:vSW}). In this regime the film departs from a glass thanks to high pressures existing during acoustic time scale in a film which causes repulsion of a film from a glass \cite{Inogamov2014nanoBump,Inogamov2015nanoBump,Inogamov2016nanoscResLett}. The second family of points corresponds to enhanced energies of a pulse. Here an edge of a hole is defined by strength of the decaying shock wave running in a substrate under a film. Now the pressure field behind the shock in a glass pushes initially motionless film toward the vacuum side. Thus in the first family the high pressures are in a film, while in the second family the pressure field acts from a glass.
\label{fig2exprmnt}
}
\end{figure}


\section{Experimental results}

Our experimental results are presented in figures~\ref{fig1exprmnt} and \ref{fig2exprmnt}. Figure~\ref{fig1exprmnt} shows the scanning electron microscope (SEM) images of the holes created by the solitary laser impact. The scale bars in the eight images all are equal to 1 micron. The images are placed from left to right and from the upper series to the lower series according to energy of a laser pulse. These energies are 48, 64, 100, 240, 320, 400, 800, and 960 nJ. We see how the radius of a hole rise as energy increases. The plot which shows dependence of square of the hole radius as function of energy is given in figure~\ref{fig2exprmnt}.

We see very definite kink in figure~\ref{fig2exprmnt}. Our paper is devoted to explanation of this kink---why it appears, to what physical phenomenon it corresponds. Position of the kink relates to energies $\sim 100$~nJ. The kink in figure~\ref{fig2exprmnt} locates inside the interval of energies corresponding to the holes, see figure~\ref{fig1exprmnt}.

In experiments the second harmonics with wavelength 515 nm was used. Duration of a pulse was 200 fs (FWHM---full width half maximum). Numerical aperture was NA=0.25. We use a single laser impact onto target. Thin silver film with thickness 50 nm was our target. Silver was deposited onto 1 mm thick silica plate.

\begin{figure}
\begin{center}
\begin{minipage}[b]{0.47\columnwidth}
\includegraphics[width=1\columnwidth]{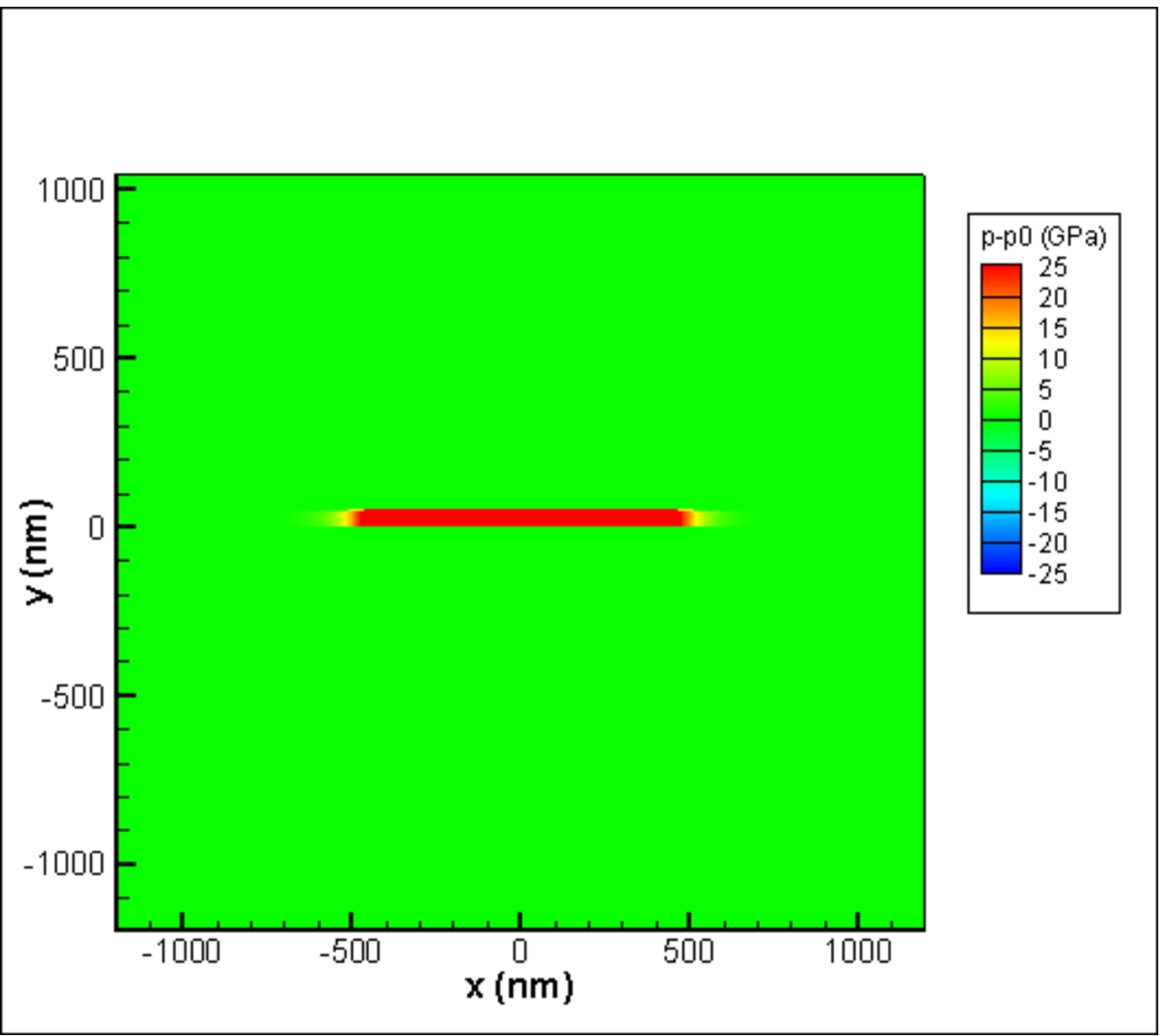}
\begin{center}\pt(a)\end{center}
\end{minipage}
\hspace{0.04\columnwidth}
\begin{minipage}[b]{0.47\columnwidth}
\includegraphics[width=1\columnwidth]{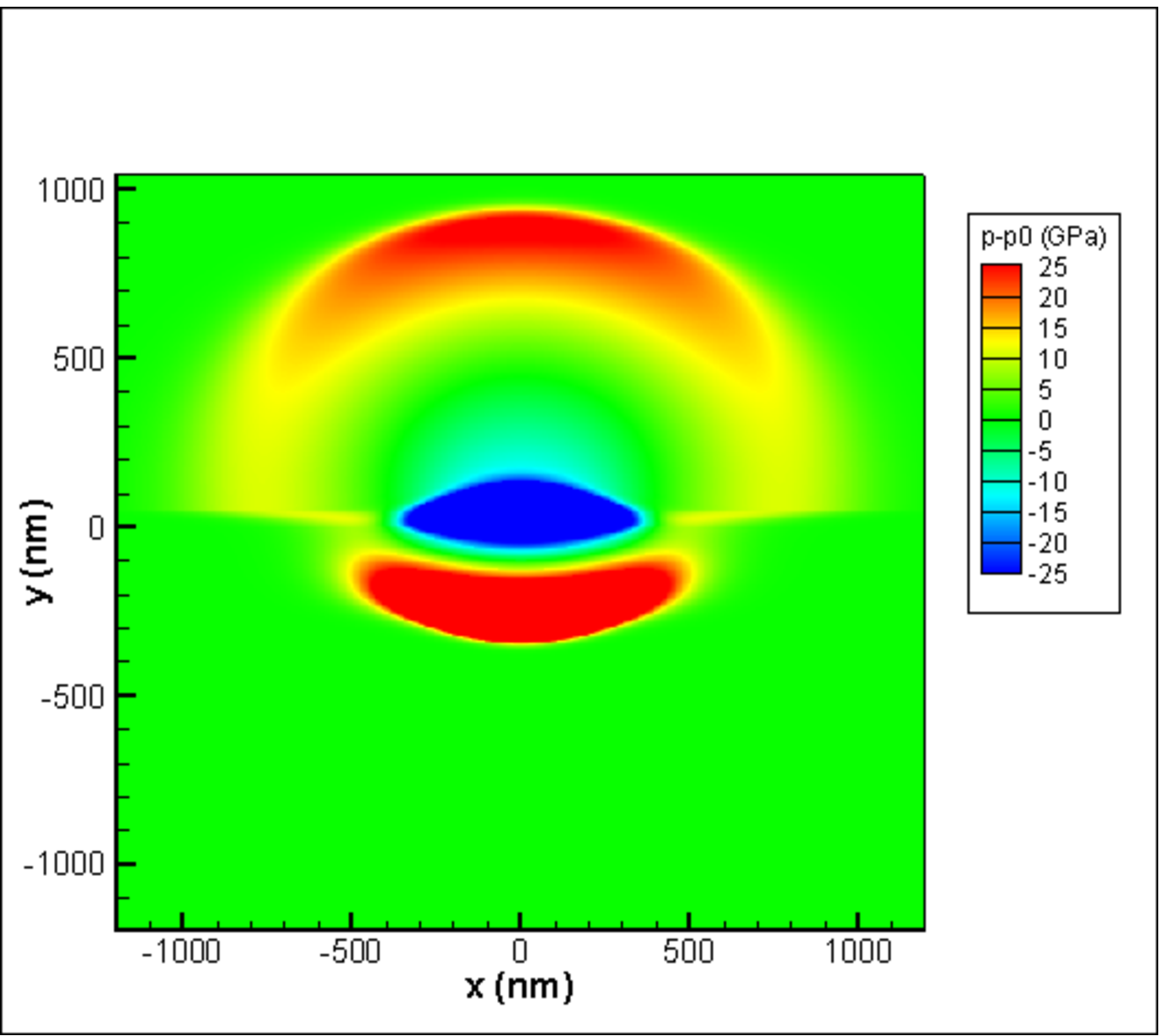}
\begin{center}\pt(b)\end{center}
\end{minipage}
\end{center}
\caption{\label{fig3} Part~\pt(a) shows the initial $t=0$ $p-p_{\rm ini}$ distribution; $p_{\rm ini}=100$~GPa. Laser energy is absorbed in gold film; heating rises pressure in a film, situation prior to film expansion is shown. Distribution of absorbed heat is given by Gaussian function with $R_{\rm L}=250$~nm, it is shown in figure~2\pt(a) as the curve 4. Part~\pt(b) shows the instant $t=20$~ps, $p-p_{\rm ini}$ distribution. Shock in substrate (red-yellow crescent with ends up) is transforming from 1D to 2D shape. There are shock plus rarefaction in ``vacuum'' (the upper semi space is the effective ``vacuum'' with low inertia weakly affecting dynamics of a film). The shock in our effective vacuum moves fast, it is weak (the light yellow crescent near the upper wall is the instant shock compressed area in the upper semi space). In the instant area occupied by the rarefaction wave the pressure $p-p_{\rm ini}$ is negative. }
\end{figure}

\section{Shock driven tearing off of a film from substrate}

We consider the process of shock pushing of a film. Temporal behavior of velocities $v_{\rm cm}(t)$, see (\ref{eq1:vSW}), for the three observation points are presented in the inset in figure~2 \pt(b). At the growing part the velocities $v_{\rm cm}(t)$ are gradually accumulated during SW acceleration of the observation point, when the shock compressed half-spherical shell passes the point and pressure under the point in figure~1\pt(b) is positive. The first dependence taken in the point $x=0$ should be omitted because it belongs to the inner part where the thermomechanical process operates. The two other observation points at $x=600$~nm and $x=1080$~nm are far apart from the heating radius $R_{\rm L}=250$~nm of the Gaussian beam in figure~2. Nevertheless they acquire significant velocities $v_{\rm cm}(t_{\rm B})$ (\ref{eq1:vSW}) equal to 360 and 240 m/s; these are the maximum values of the functions $v_{\rm cm}(t)$ in figure~2\pt(b). These velocities are enough to separate a film from substrate \cite{Wang2017}.



\begin{figure}
\begin{center}
\begin{minipage}[b]{0.47\columnwidth}
\includegraphics[width=1\columnwidth]{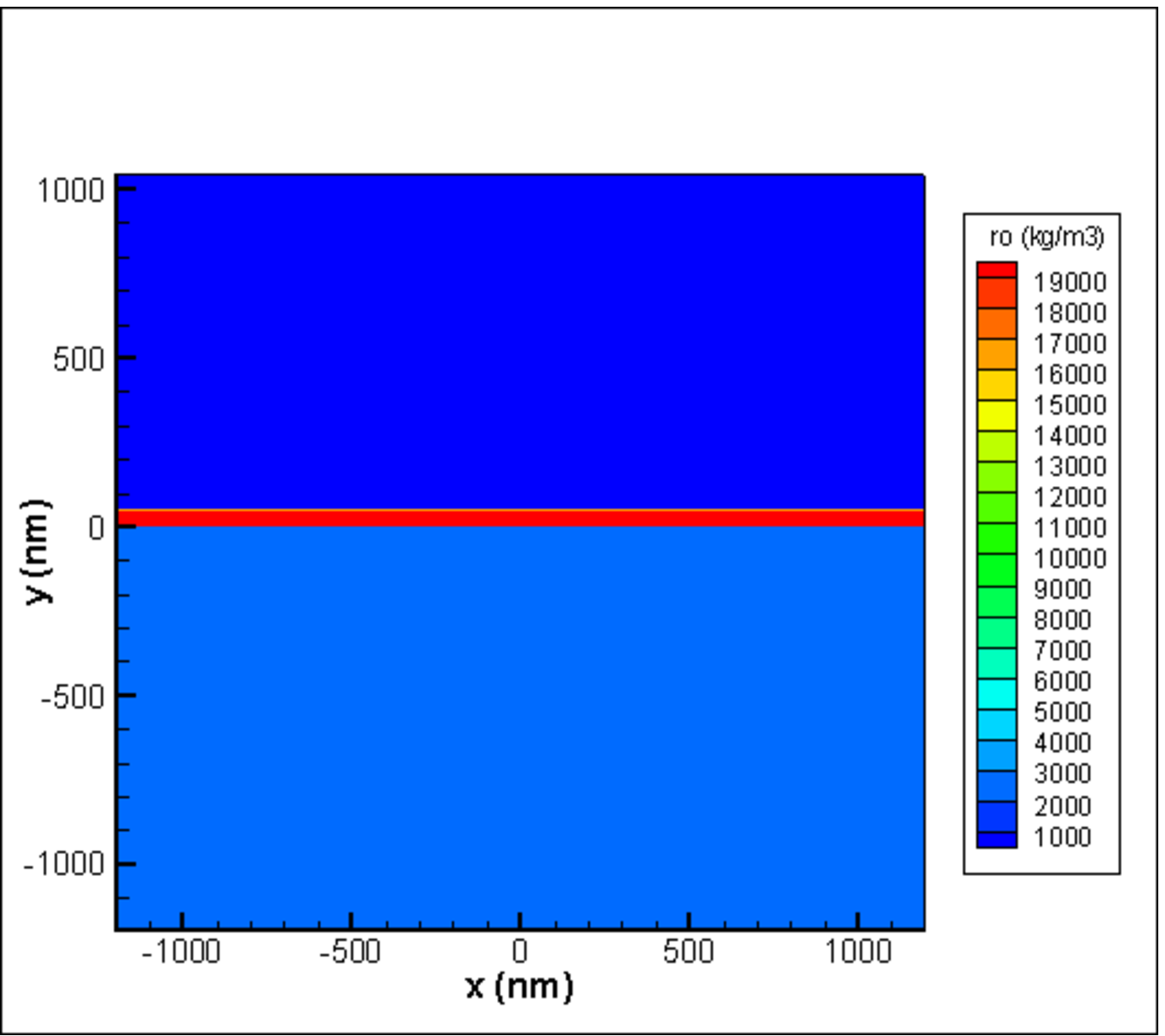}
\begin{center}\pt(a)\end{center}
\end{minipage}
\hspace{0.04\columnwidth}
\begin{minipage}[b]{0.47\columnwidth}
\includegraphics[width=1\columnwidth]{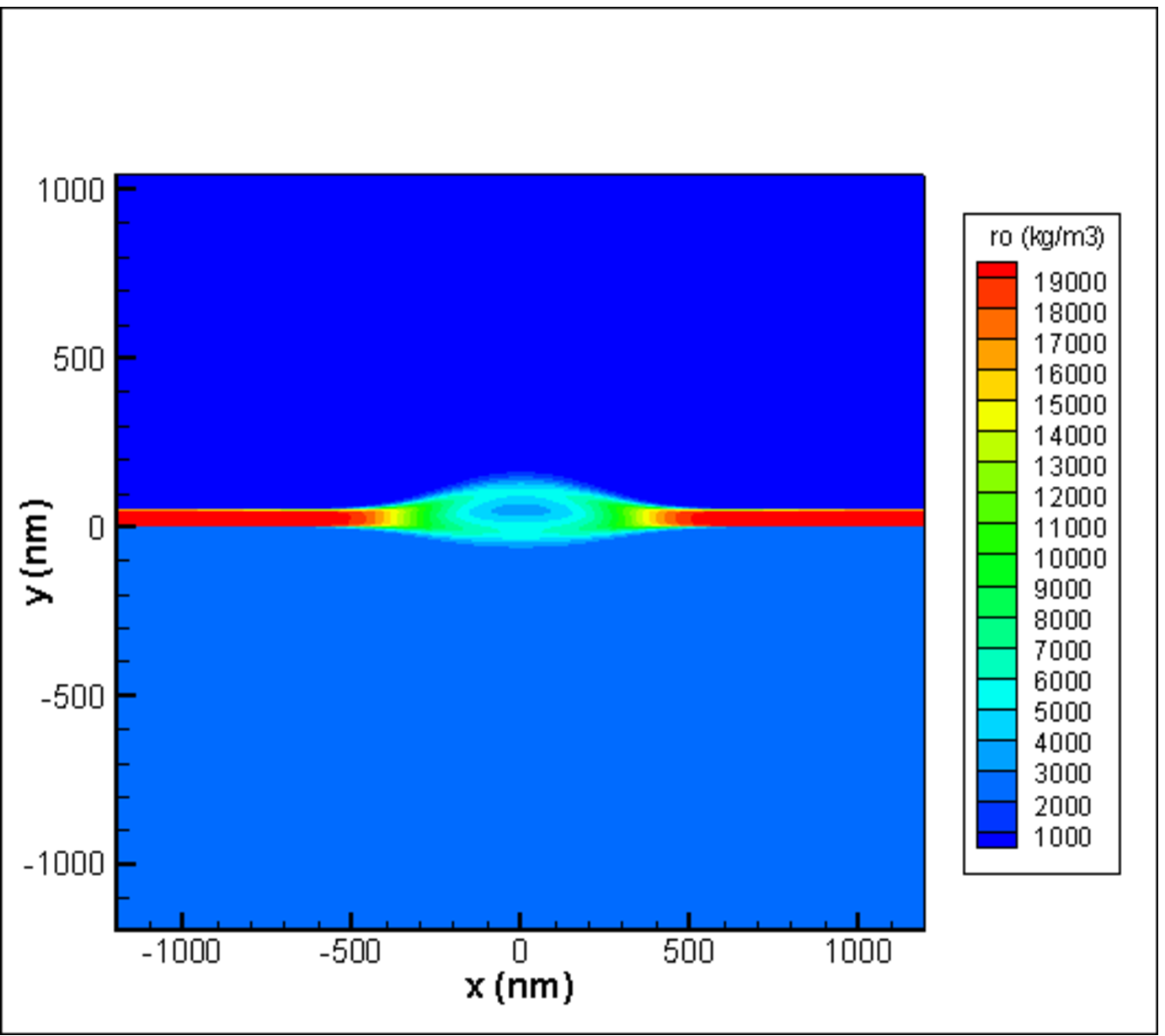}
\begin{center}\pt(b)\end{center}
\end{minipage}
\end{center}
\caption {\label{fig4} Part~\pt(a) shows the initial $t=0$ density distribution. This is the snapshot prior to onset of motion, therefore the sharp rise of pressure shown in figure~\ref{fig3}\pt(a) does not yet affect density field. From up to bottom: vacuum (deep blue, small density outside the ladder of colors), film (dense gold, the red strip in the middle), and glass substrate (the bottom blue rectangular, initial density is $\rho_{\rm sbstr}=2300$~kg/m$\!^3)$ are shown. Part~\pt(b) corresponds to the instant $t=20$~ps, density distribution. We see swelling of gold film and compression in glass. The compression is the weak blue half-sphere under the swelling. The compression corresponds to the region of enhanced pressure in the shape of crescent well seen in figure~\ref{fig3}\pt(b). This is the region of the shock compressed silica. It follows the shock in glass. }
\end{figure}


Thus we come to the main conclusion of the paper. The regime with shock driven separation of a film exists.

To calculate velocities shown in figure~2\pt(b) we put the pressure dependencies $p(x_i, y=-40, t)$ into integral (\ref{eq1:vSW}); here $x_i=0, 600$ and 1080~nm, see the inset in figure~2 \pt(b); this inset corresponds to the right part of region near and slightly below the film shown in the computational box in figure~2\pt(a). We use the level $y=-40$~nm in glass (below the moving film) for calculation of pressure acting on a film because a gold film is dense and shifts slowly thus this level is always inside glass. Pressure gradients are sharp in a film (Au is dense and a film is thin) and small inside the shock compressed area in silica (because shock propagates large distances relative to the thickness of a film). Therefore pressures at this horizontal level are approximately equal to the pressure in the observation point at the bottom boundary of a film, indeed $|y=-40 \, {\rm~nm}|\ll x_i,$ $i=2, 3$---film is thin, distance between the level $y=-40$~nm and the film is small relative to propagation distance $$r=\sqrt{x^2+y^2}$$ of the shock; this distance is a measure of gradients $\propto 1/r$ behind SW.

\section{Simulation details and results}


It is interesting that there are many computational works devoted to laser action onto films, but all of them are or about MD
\cite{Ivanov2013,Ivanov2017,Rouleau2014,SHIH20173,
Inogamov2014nanoBump,Inogamov2015nanoBump,Inogamov2016nanoscResLett,AnisimovFLAMN:2017,Wang2017} simulations, or for 1D hydrodynamic modeling \cite{Inogamov2014nanoBump,Inogamov2015nanoBump,Wang2017, InogamovJETPLett:1999}--the 2D hydrodynamic works are absent if we exclude papers \cite{Meshcheryakov2005,Meshcheryakov2013}. This means that 2D hydrodynamic finite difference approximation of ablation problem is difficult.



\begin{figure}
\begin{center}
\begin{minipage}[b]{0.47\columnwidth}
\includegraphics[width=1\columnwidth]{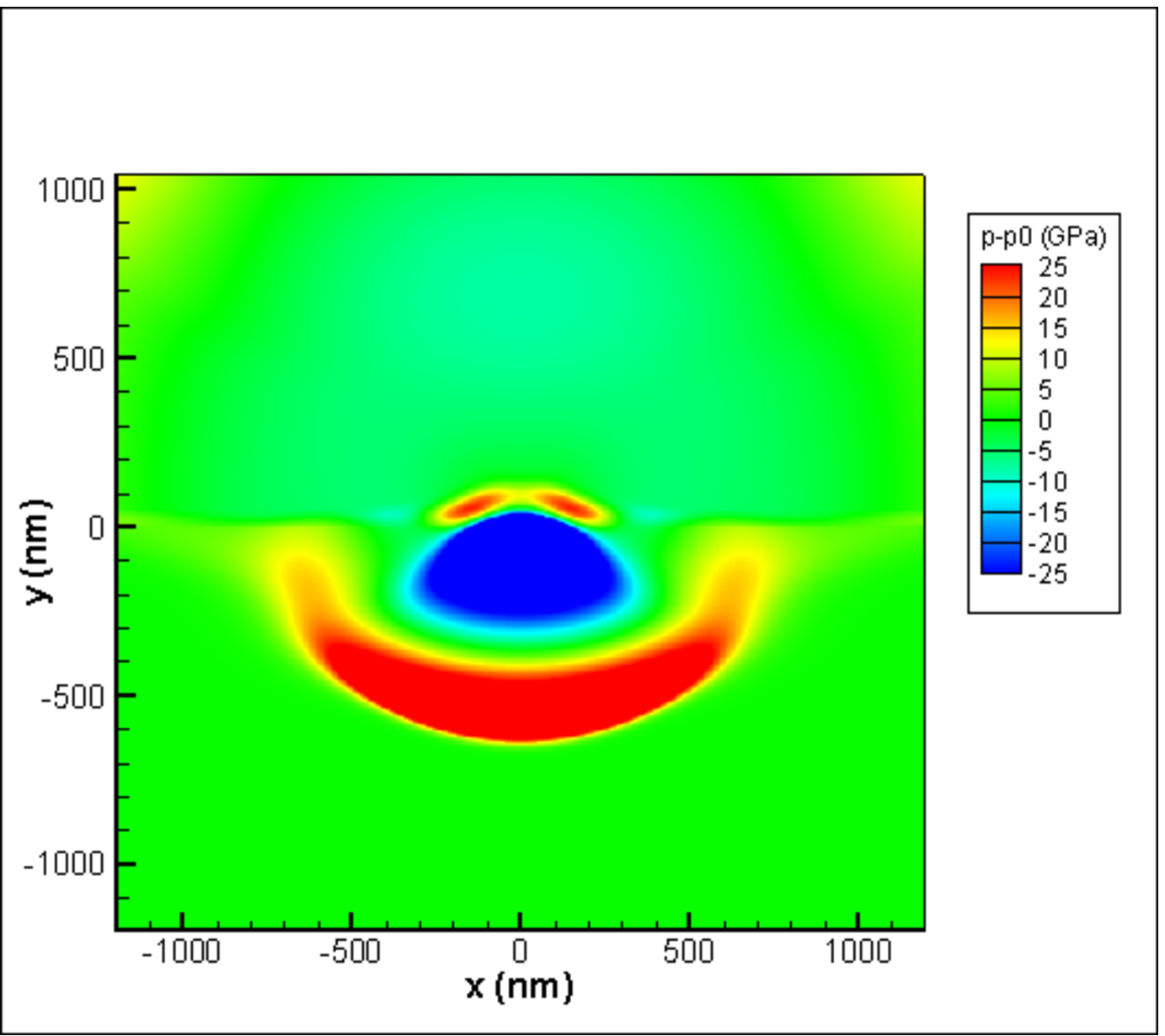}
\begin{center}\pt(a)\end{center}
\end{minipage}
\hspace{0.04\columnwidth}
\begin{minipage}[b]{0.47\columnwidth}
\includegraphics[width=1\columnwidth]{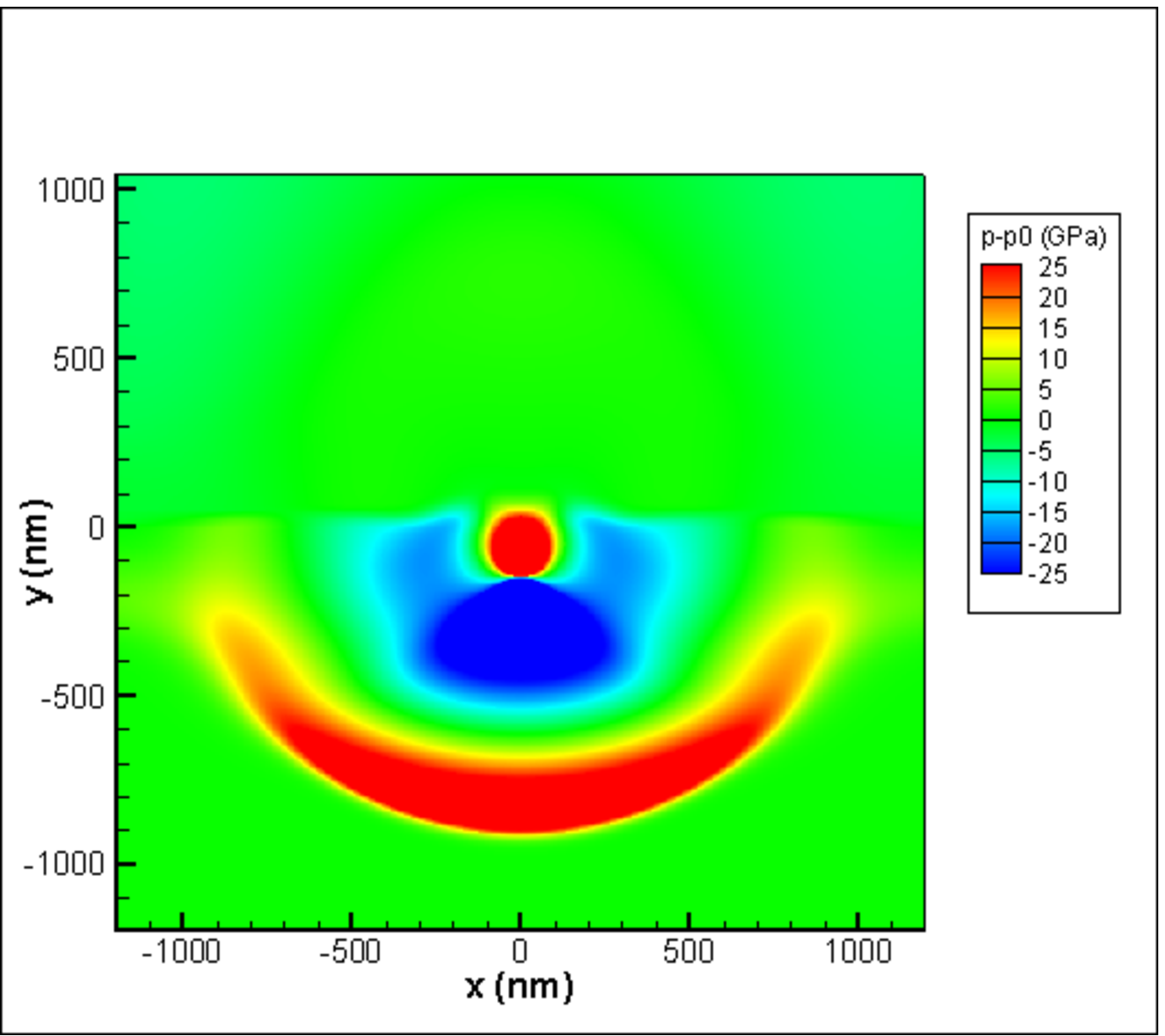}
\begin{center}\pt(b)\end{center}
\end{minipage}
\end{center}
\caption {\label{fig5} Part~\pt(a) shows the instant $t=40$~ps, $p-p_{\rm ini}$ distribution. Shock in vacuum passes away, only weak yellow traces of small compressions are seen in the upper left and right corners. Also weak rarefaction below zero level of $p-p_{\rm ini}$ in vacuum is seen---this is the weak blue area above the laser illuminated spot. Important for our model of film acceleration by shock compressed half-sphere moving downward and along the film--glass contact is the bright yellow-red large crescent with the ends up in the bottom rectangular corresponding to glass substrate. The ends of the crescent propagates along the film, positive pressure $p-p_{\rm ini}$ in the ends acts to accelerate a film into vacuum side. There is the deep blue semishpere of rarefaction below the zero level in glass. But this negative pressure $p-p_{\rm ini}$ in this rarefaction cannot turn back a film if it is separated from glass before. We should not pay attention to the compression near the central area of the film (this area corresponds to the laser focal spot). This compression appears thanks to oscillation of the film disrupted near its middle plane and after disruption expanded in to two sides: the upper part expands into vacuum side, while the bottom part---into the glass side. Effective pressure $p_{\rm ini}=100$~GPa in vacuum and in glass turns these parts back (this pressure supports initial static state). The parts collide and this collision induces the second rise of pressure. In real life vacuum cannot return back the upper part of a film belonging to a film inside heating radius $R_{\rm L},$ and this part moves away. Part~\pt(b) shows the instant $t=60$~ps, $p-p_{\rm ini}$ distribution. The main features here are the compressed half-sphere shell and the rarefaction region in glass.
We do not pay attention to the new cycle of compression near the central area of a film because, as was said in caption to figure~5\pt(a), it is consequence of the background pressure of our effective vacuum which turns back of the upper part of the gold film. In reality this part freely moves away. }
\end{figure}




\begin{figure}
\begin{center}
\begin{minipage}[b]{0.47\columnwidth}
\includegraphics[width=1\columnwidth]{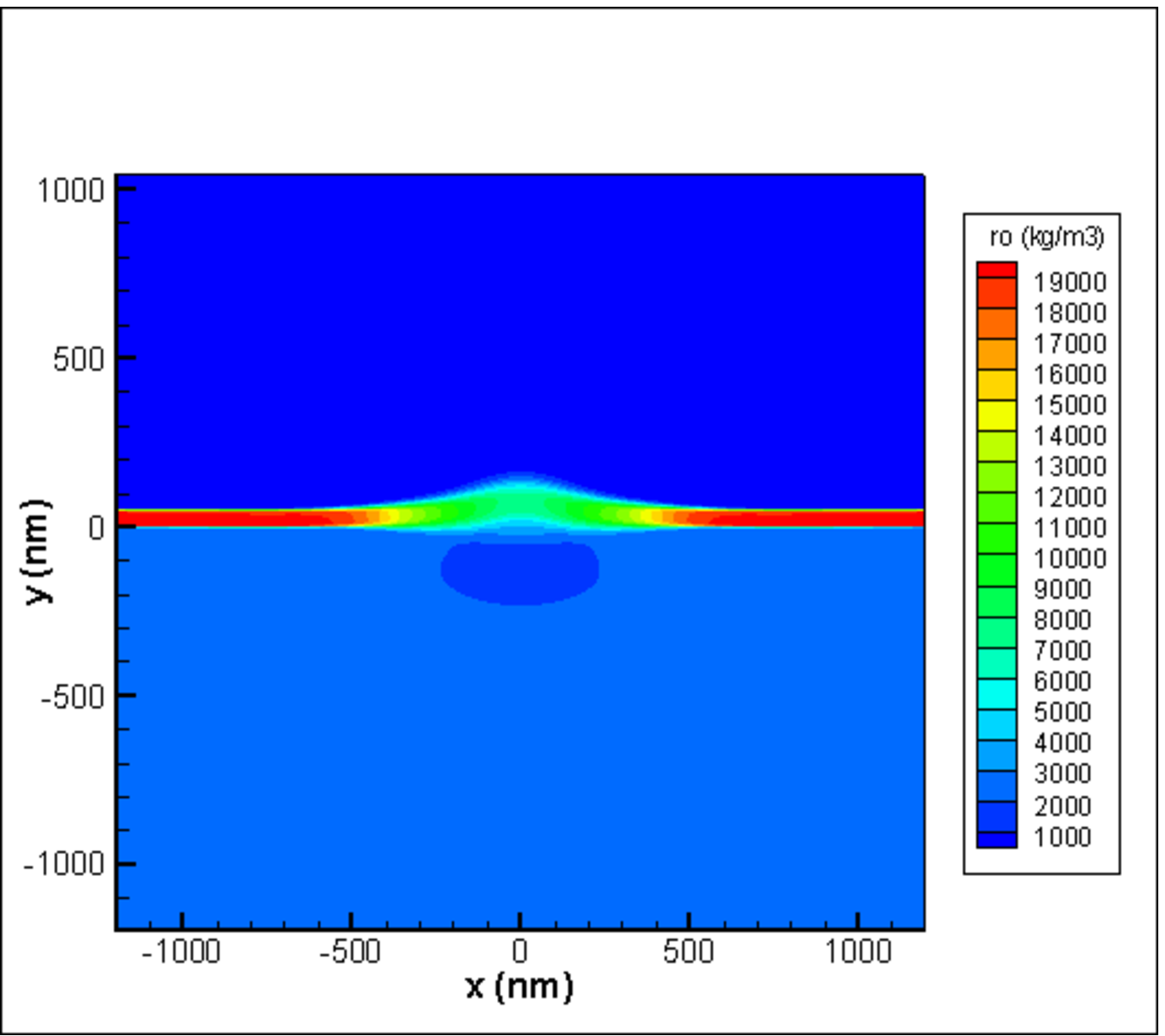}
\begin{center}\pt(a)\end{center}
\end{minipage}
\hspace{0.04\columnwidth}
\begin{minipage}[b]{0.47\columnwidth}
\includegraphics[width=1\columnwidth]{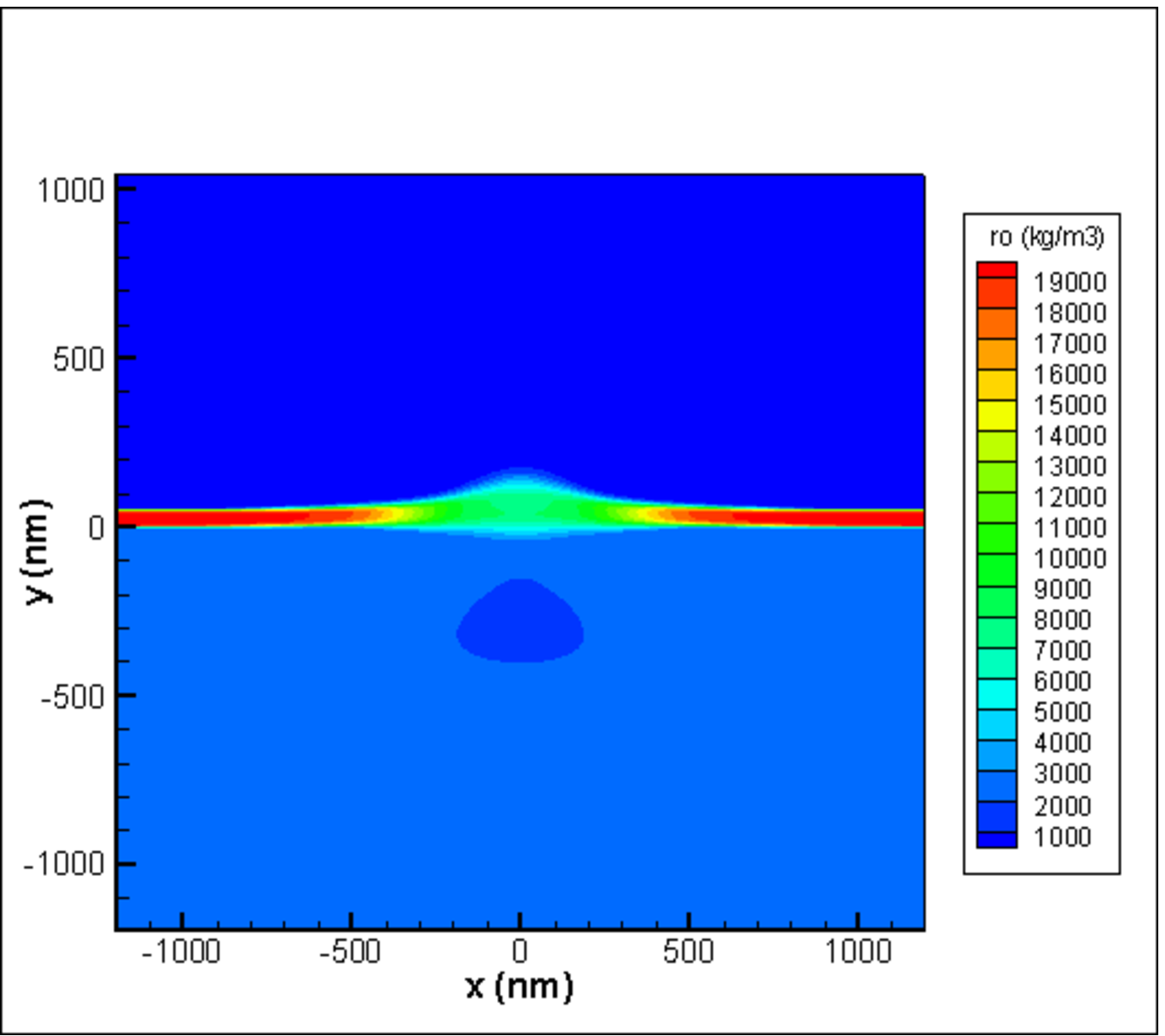}
\begin{center}\pt(b)\end{center}
\end{minipage}
\end{center}
\caption { \label{fig6} Part~\pt(a) shows the instant $t=40$~ps, density distribution.
We see compression followed by rarefaction in the substrate and swelling of gold film. Part~\pt(b) shows the density field at the instant $t=60$~ps.
Compression half-sphere and rarefaction region in glass are seen. }
\end{figure}

\begin{figure}
\begin{center}
\begin{minipage}[b]{0.47\columnwidth}
\includegraphics[width=1\columnwidth]{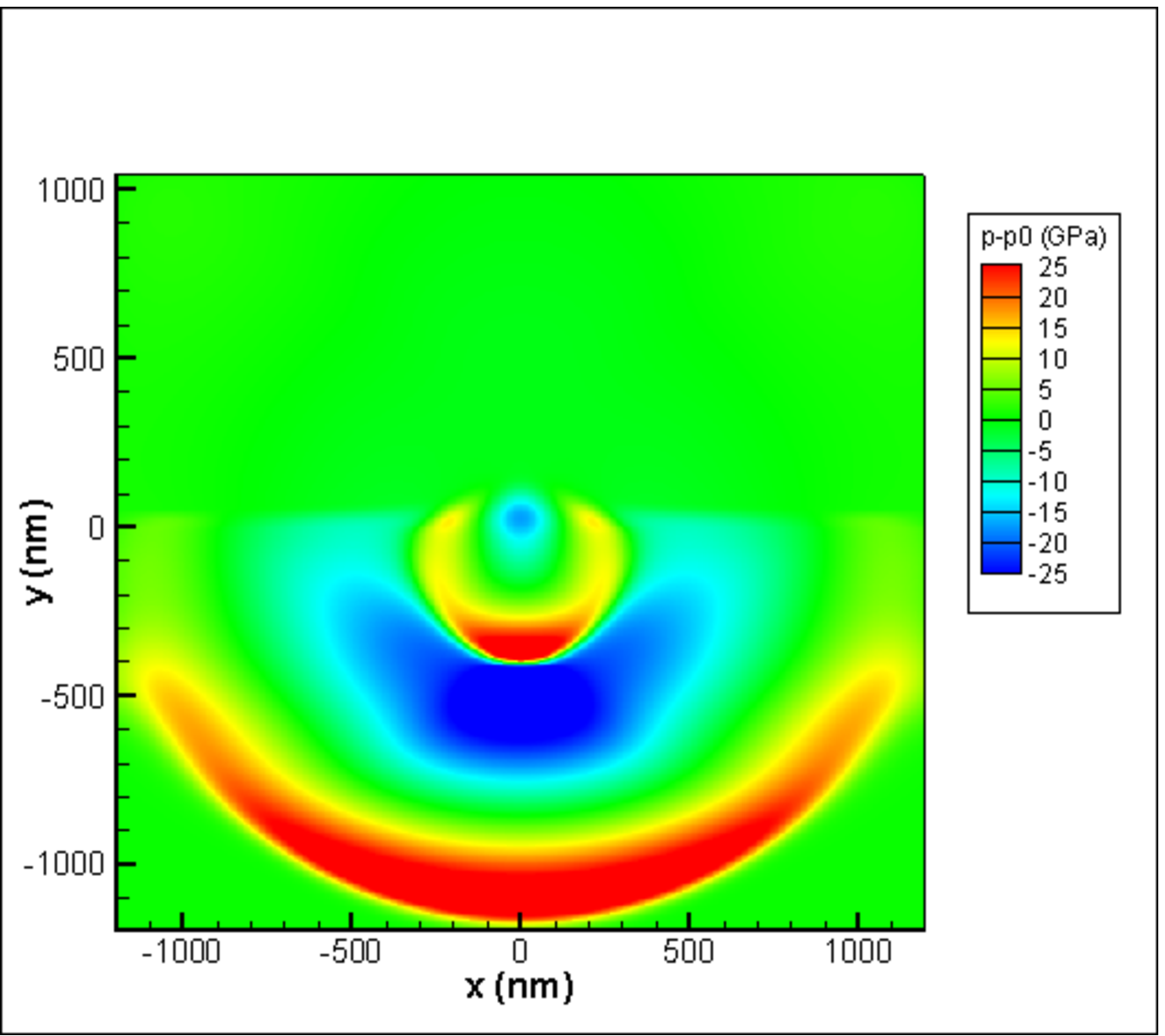}
\begin{center}\pt(a)\end{center}
\end{minipage}
\hspace{0.04\columnwidth}
\begin{minipage}[b]{0.47\columnwidth}
\includegraphics[width=1\columnwidth]{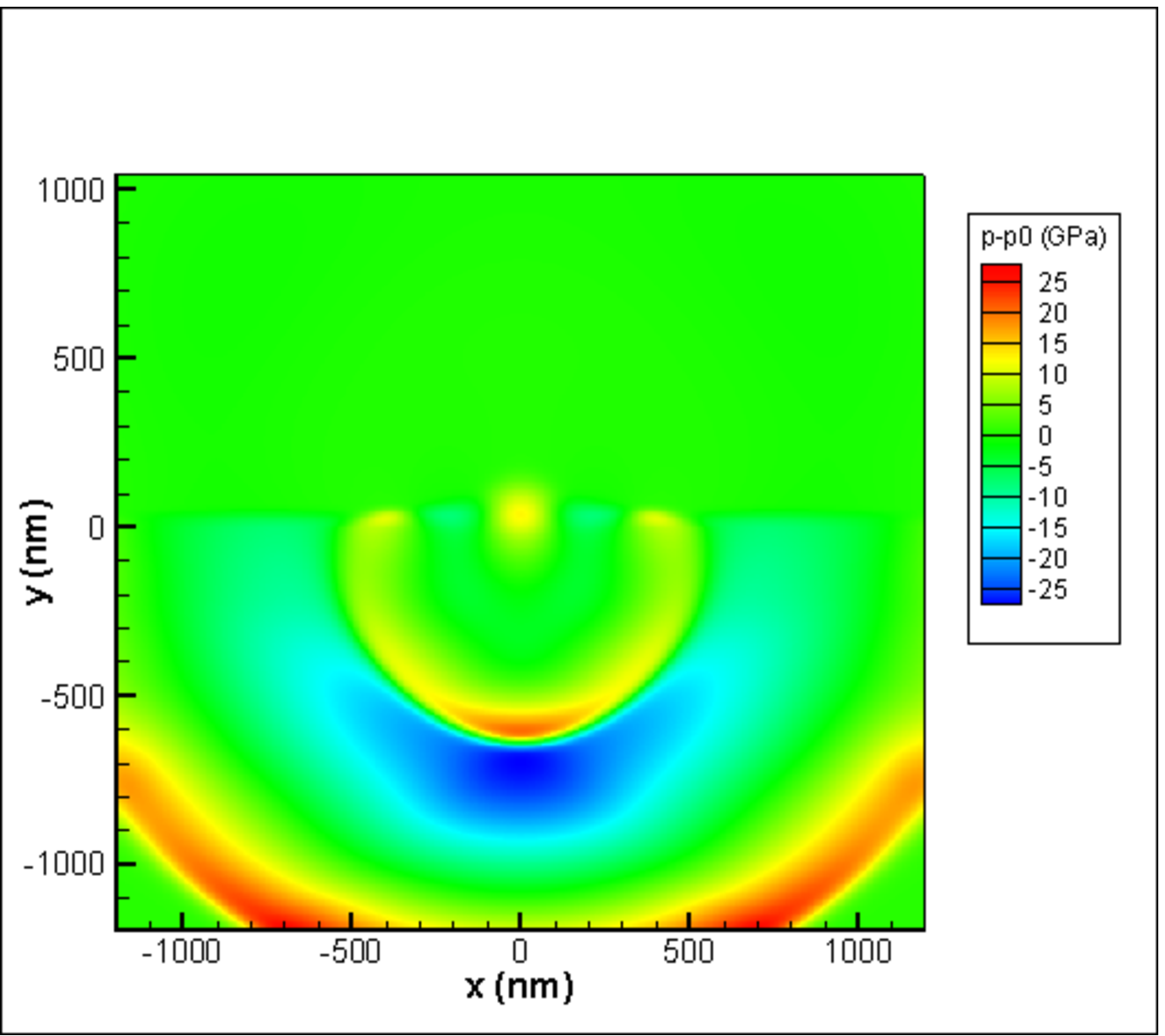}
\begin{center}\pt(b)\end{center}
\end{minipage}
\end{center}
\caption {\label{fig7} Part~\pt(a) shows the instant $t=80$~ps, $p-p_{\rm ini}$ distribution. Shock in glass almost achieves the bottom wall with free streaming boundary condition. The acceleration of the film by the compression half-sphere shell (crescent) in glass sliding along the film is finished inside our computational box because the half-sphere shell leaves the box through the lateral walls of the box. The new parasitic rarefaction appears near the center of heating due to secondary oscillations of the broken expanding parts of the film which further return and collide. Part~\pt(b) shows the instant $t=100$~ps, $p-p_{\rm ini}$ distribution. Shock in glass leaves the box through transparent walls. }
\end{figure}
\begin{figure}
\begin{center}
\begin{minipage}[b]{0.47\columnwidth}
\includegraphics[width=1\columnwidth]{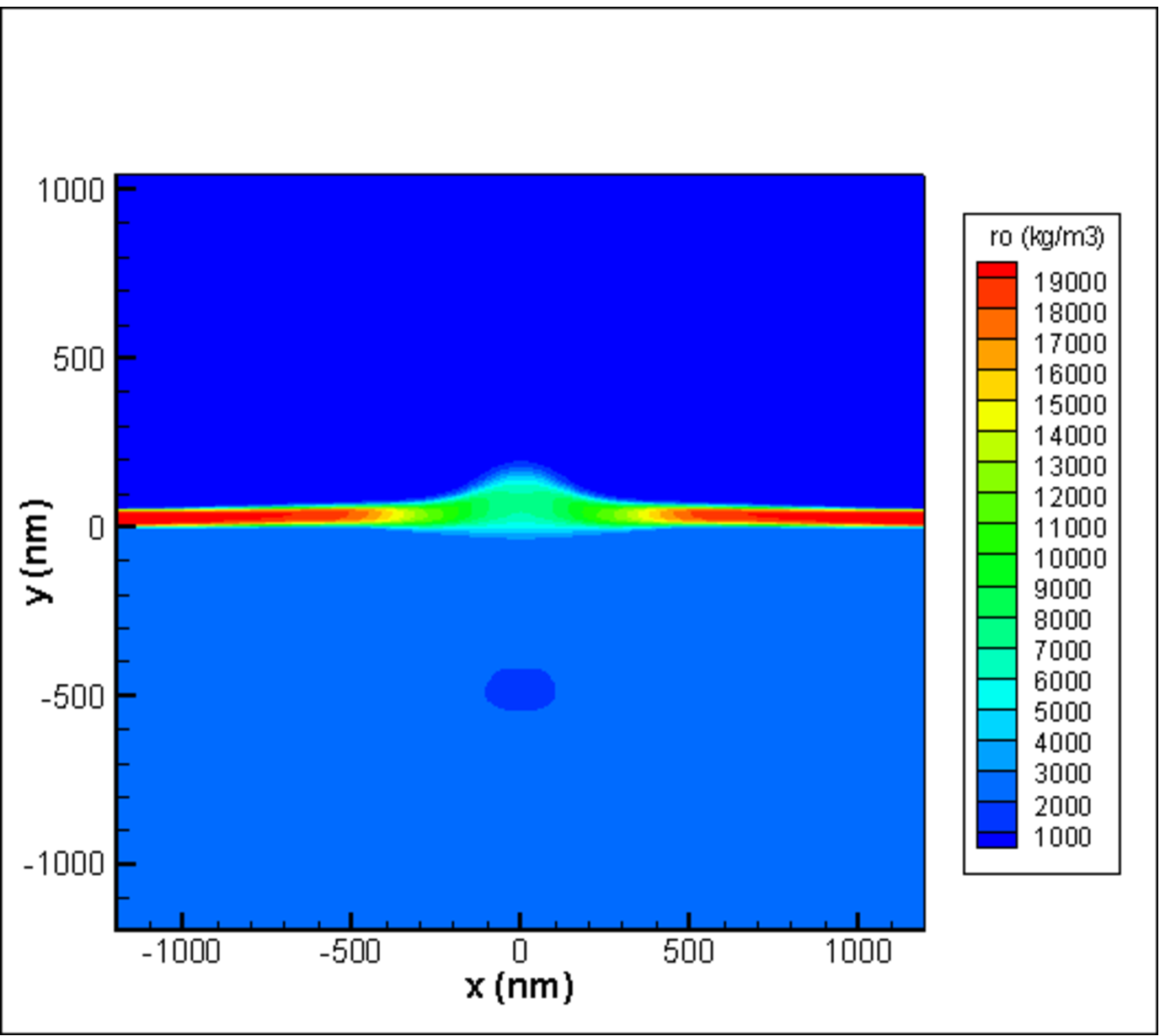}
\begin{center}\pt(a)\end{center}
\end{minipage}
\hspace{0.04\columnwidth}
\begin{minipage}[b]{0.47\columnwidth}
\includegraphics[width=1\columnwidth]{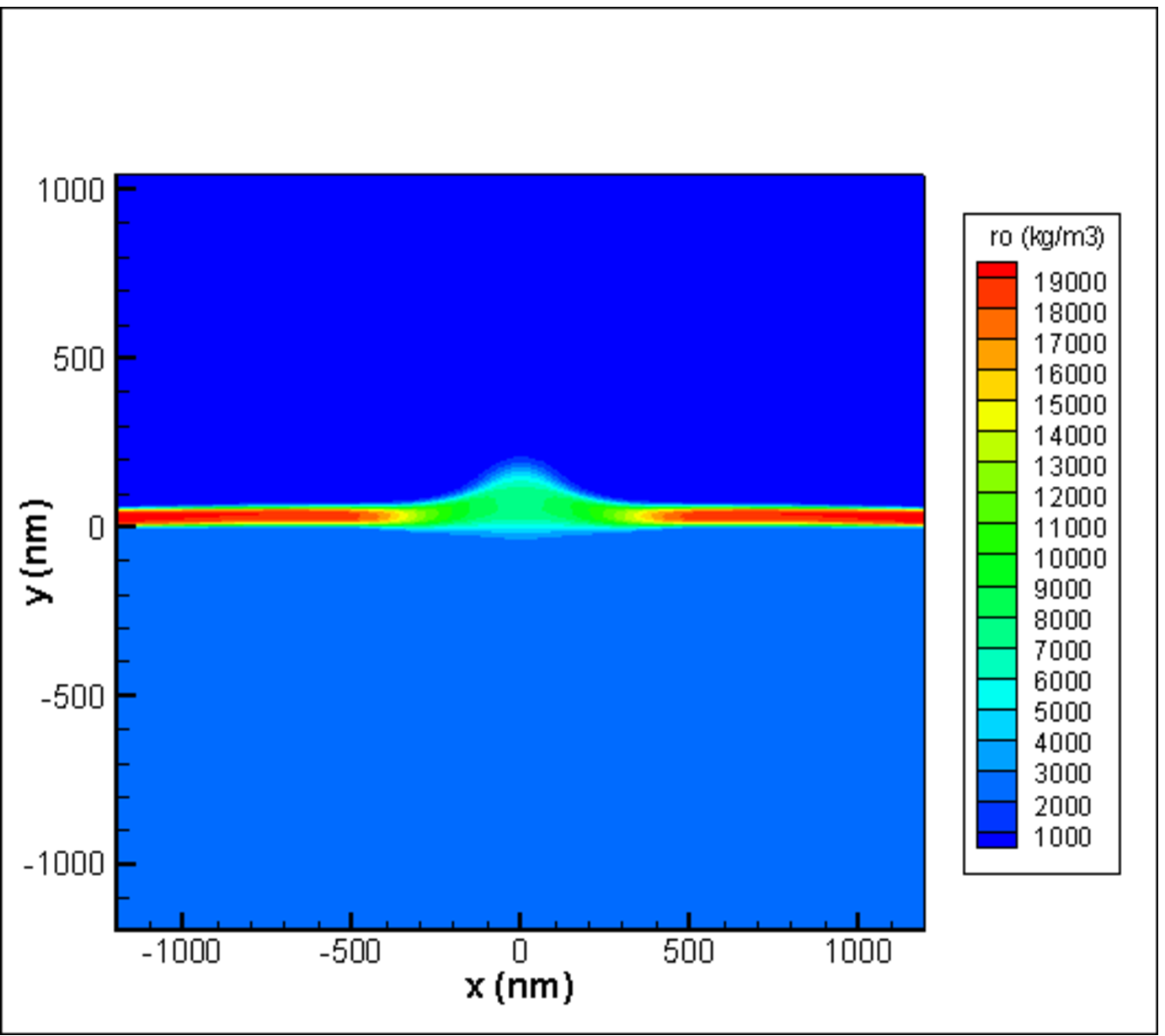}
\begin{center}\pt(b)\end{center}
\end{minipage}
\end{center}
\caption {\label{fig8} Part~\pt(a) shows the instant $t=80$~ps, density distribution.
We see compression followed by rarefaction in the substrate and swelling of gold film. Part~\pt(b) shows the instant $t=100$~ps, density distribution.
We see compression followed by rarefaction and new compression (the parasitic one) in the substrate and swelling of the gold film. }
\end{figure}

Results of our largest in spatiotemporal scales simulation are presented in movies posted at youtube
\cite{density,p125,p110}
and figures~\ref{fig3}--\ref{fig8}. These results are used to plot velocity dependencies given in figure~2\pt(b). Figures~\ref{fig3}--\ref{fig8} and movies show formation and expansion of two shocks: one in silica while second in ``vacuum''. In simulation we cannot drop down density in vacuum semi space to zero. Thus some small density 200~kg/m$\!^3$ for ``vacuum'' is necessary; small relative to initial densities of gold 19300~kg/m$\!^3$ and silica 2300~kg/m$\!^3.$ We use ideal gas approximation with adiabatic exponent $\gamma=5/3$ for equation of state. Of course, behavior of ideal gas seems is far from behavior of condensed media. The main difference is in rigidity of condensed matter at small temperatures versus softness of ideal gas. Typical bulk moduli for condensed matter are $\sim 100$~GPa. Adiabatic bulk modulus for ideal gas is $B=\rho (\partial/\partial\rho) p|_s = \gamma p.$

To compensate difference in moduli $B$ between condensed matter and gas we impose large initial temperatures in gas. Initially three layers of vacuum 3, film 1, and silica 2, see figure~2\pt(a), have equal pressures and therefore they are in dynamic equilibrium. Initial pressure in these three layers is 100~GPa. If during motion pressure $p_g$ in gas drops below $p_{\rm ini}=100$~GPa $(p_g < p_{\rm ini}),$ this will mean that in the corresponding condensed matter pressure drops below zero value to negative value equal to minus $|p_{\rm ini}-p_g|.$

To achieve equal pressures at different densities we have to impose different temperatures. They are in kelvins [K]: $4.22705 \times 10^6$ $(\approx 4200$~kK) for vacuum (rectangular~3 in figure~2\pt(a), 43804 $(\approx 44$~kK) for film (narrow rectangular~1), and 367570 $(\approx 370$~kK) for silica (rectangular~2). The same gas with atomic weight 70 fills all three rectangular.
We take the dependence of hole radius $R_{\rm h}$ (in 50 nm Au film) on pulse energy $E$ from paper \cite{Wang2017}, see figure~1(h) in this paper.
Let absorbed fluence is
$F_{\rm abs} = 2.4$~J/cm$\!^2;$ simulations presented here and figure~2 \pt(b) correspond to this value of $F_{\rm abs}.$ Then according to the dependence from \cite{Wang2017} the radius $R_{\rm h}$ is 4--5~microns---approximately ten (!) times more than the laser beam radius $R_{\rm L}.$ Separation velocities of gold film from glass substrate are just ten or few tens of~m/s, see \cite{Wang2017}. Threshold for separation is low because adhesion between Au and glass is weak. Velocities $v_{\rm cm}(t_{\rm B})$ shown in figure~2\pt(b) are large, thus the distance $x_{\rm sw}$ of separation under action of SW is more long than the horizontal width 1.2~$\mu$m of the half of computational box in figures~2\pt(a) and 2\pt(b) (see inset).

We cannot directly compare the dependence from figure~1(h) \cite{Wang2017} and our results on $x_{\rm sw}(F_{\rm abs}).$
First, the width of our computational box in figure~2\pt(a) is not enough large to follow decrease of velocity $v_{\rm cm}(t_{\rm B})$ at large distances $x.$
Second, many runs are necessary to plot dependence $x_{\rm sw}(F_{\rm abs}).$
Third, our simulation is in 2D$_{\rm p}$ plane geometry (Cartesian coordinates $x,y)$ while in \cite{Wang2017} geometry is axisymmetric 2D$_{\rm a}$: cylindrical radius $r_{\rm cyl}$ and height $z$ along the symmetry axis are cylindrical coordinates. Thus in \cite{Wang2017} the 2D$_{\rm a}$ shock in substrate decays faster with distance than in our 2D$_{\rm p}$ case; then the 2D$_{\rm a}$ shock driven hole is smaller. In linear acoustic approximation an amplitude of the 2D$_{\rm a}$ shock decays $\propto 1/r^2$ with the propagation path $r,$ while in the case of the 2D$_{\rm p}$ shock the decay law is $\propto 1/r.$

Nevertheless here for the first time the SW driven mechanism of opening of a hole is proposed. It is alternative to the thermomechanical mechanism. New mechanism proves existence of large holes with radius $R_h$ much more than the beam radius $R_L.$ Large holes appear at large laser energies but these energies are much less than a huge value $F_{\rm tm} \exp(R_h^2/R_L^2)$ following from laser thermomechanics.



Let us calculate increase of temperature in Au film after absorption of ultrashort pulse. Temperature distribution immediately after absorption is given by Gaussian function. Increase of temperature above initial temperature $\approx 44$~kK in Au is
\begin{equation} \label{eq2:Trise}
\Delta T(x, t=0) = T_c \exp(-x^2/R_L^2),
\end{equation}
where $T_c$ is increase of temperature in the central point of a hot spot, $R_L,$ as was said above, see figure~2\pt(a), is 250~nm.
For absorbed central fluence $F_{\rm abs}=2.4$~J/cm$\!^2$ the value $T_c$ in (\ref{eq2:Trise}) is 400~kK; heat capacity of gas per unit of volume is $(3/2) n k_B,$ where $n$ is concentration of atoms. Shortly after pulse $(\tau_L=230$~fs$\, \ll t_s$ \cite{Wang2017}) pressure in the central point of a film is 10~Mbar---ten times more than initial pressure giving estimate of bulk modulus for gas.

Sharp (faster than acoustic time scale $t_s)$ increase of temperature (\ref{eq2:Trise}) triggers expansion of gold inside a hot spot and generates shock in substrate, see figure~1\pt(a). There is also another shock going up into vacuum, because vacuum in our simulation has small (1\% relative to Au) but finite density. This upper shock was not shown in figures~1\pt(a) and 1\pt(b). Both shocks are well seen in movies
\cite{density,p125,p110}
and figures~\ref{fig3}--\ref{fig8}. They form two expanding half-spherical shells above and below a film. The shells are divided by the film well seen in the movie \cite{density} showing evolution of density. The upper shell has smaller pressure amplitude relative to the SW in substrate because density of vacuum is much less than density of silica, while expansion velocities of the gold-vacuum and gold-silica contacts are approximately the same since equal pressures 100~GPa oppose expansion of Au from the vacuum and silica sides; expansion of a contact works as a moving piston initiating shock ahead the piston.

Adiabatic sound speed in our effective vacuum $c_{\rm s}|_{\rm vac}$ is approximately three times over sound speed in glass $c_{\rm s}|_{\rm sbstr},$ because $(c_{\rm s}|_{\rm vac}) \, / \, (c_{\rm s}|_{\rm sbstr}) = \sqrt{T_{\rm vac}/T_{\rm sbstr}} \approx \sqrt{4200/370}.$ Therefore SW in vacuum moves significantly faster than SW in substrate. It quickly crosses the upper rectangular~3 in figure~2\pt(a) and disappears thanks to non-reflecting boundary conditions imposed on the upper and lateral walls of rectangular. Dynamically influence of the vacuum SW onto dynamics of a gold film is very weak. A film is dense, its velocities are small thus there is small shift of the film during the passage time of SW in silica through the bottom rectangular~2 in figure~2\pt(a). We do not take into account pressure from the vacuum side in our calculations of velocities $v_{\rm cm}(t)$ shown in figure~2 \pt(b).
But, of course, our effective vacuum influences thermomechanically driven expansion of the upper subfilm into vacuum side inside hot spot radius $\sim R_{\rm L},$ see figure~1\pt(b). Thus we cannot obtain results similar to \cite{Ivanov2013,Domke:12,Rouleau2014,Inogamov2014nanoBump,Inogamov2015nanoBump,Inogamov2016nanoscResLett,AnisimovFLAMN:2017,Wang2017} for expansion of gold shell in vacuum. In some sense our situation with the upper Au shell and effective vacuum resembles situations studied in \cite{Ivanov2017,SHIH20173} with expansion of a film into surrounding liquid. But our task in this paper is to understand and estimate influence of the under-film SW onto dynamics of a film in the outside (relative to $R_{\rm L})$ region.



\section{Advantages and shortcomings of gaseous approximation}

We take finite initial pressure $p_{\rm ini}$ instead of $p_{\rm ini}=0.$ In case of condensed matter in its initial state prior to laser impact we have $p_{\rm ini}=0.$ In condensed state an adiabatic dependence $p(\rho, s)$ passes a binodal (bin) line with $p=0$ (we neglect vapor pressure) at finite densities $\rho_{\rm bin}(s);$ here $s$ is the entropy. By putting the level of initial pressure in our gas $p_{\rm ini}=100$~GPa at normal densities of gold and silica we semiquantitatively correspond to condensed matter. In our gaseous approximation with finite background $p_{\rm ini}=100$~GPa pressures below 100 GPa relates to the negative pressures in the condensed case. There is difference connected with slope of our adiabatic dependence $p=s \rho^\gamma$ in the intersection point $p_{\rm ini}$ corresponding to dynamic equilibrium of our gaseous target from substrate, film, and effective vacuum. Speed of sound in our effective silica is $c_{\rm s}|_{\rm sbstr}=\sqrt{\gamma k_{\rm B} T_{\rm sbstr}/M}=8.5$~km/s at presented above values for $M$ and $T_{\rm sbstr}.$ This value is approximately twice higher than $c_{\rm s}$ in real silica.

Another semiquantitative disadvantage is too deep negative $(p-p_{\rm ini})$ pressure $p$ well in gaseous approximation. Since we take $p_{\rm ini}=100$~GPa then to spall a gold film we have to decrease pressure down to $p=0.$ If condensed matter breaks off at extensions no more than 1.5 times relative to equilibrium density (thus at finite density, the $p$-well has a minimum at this density) then our gas breaks off at density equal to zero. Therefore we cannot correctly describe spallation in the gas approximation.
But for the problem solved in this paper the correct description of stretching at large negative pressures $(p-p_{\rm ini})$ is not important. Only positive part of pressure $(p-p_{\rm ini})$ and the travelling with speed of sound edge of the half-spherical shell where $p-p_{\rm ini}=0$ are important for calculation of accumulated velocity according to expression~(\ref{eq0:vSW}). This is the main advantage of our scheme of solution. This gives us opportunity to describe necessary for our scheme stage of loading of the Au-film from its bottom boundary by the half-spherical shocked shell sliding with speed of sound along the film.

The sequence of events are shown in movies \cite{density,p125,p110} and figures~\ref{fig3}--\ref{fig8}.
In movies full pressure $p$ is presented while in figures~\ref{fig3}--\ref{fig8} we show deviation from background $p-p_{\rm ini}.$

\section{Conclusion}
Above we compare the thermomechanical kick-off a film inside its laser heated spot from the one side versus the kick connected with a shock propagating inside substrate matter under a film from the another side. In previous literature only the first case was investigated. It is shown that above a definite threshold a diameter of a final hole in a film is defined namely by the rate of decay of a shock under film. This new result allows explaining appearance of large holes in experiments presented in the paper. Diameters of these large holes significantly overcome diameter of laser beam at a target surface.

\ack
Authors (INA, physical model) acknowledge the financial support from the Federal Agency for Scientific Organizations of the Russian Federation (FASO RF) for the ITP RAS (project No.\,0033-2018-0004) and (VVS, numerical simulation) acknowledge the financial support from the FASO RF for the ICAD RAS. Authors (INA, VAK) also acknowledge support from the Russian Foundation for Basic Research (grant No.\,16-08-01181) used for development of theoretical background.

\section*{References}
\bibliographystyle{iopart-num}
\bibliography{Article_256}

\end{document}